\documentstyle[prl,aps,epsfig,color]{revtex}
\newcommand{\commento}[1]{}
\newcommand{\be}{\begin{equation}}
\newcommand{\ee}{\end{equation}}
\newcommand{\beq}{\begin{eqnarray}}
\newcommand{\eeq}{\end{eqnarray}}

\newcommand{\cu}[0]{\mathrm{CuO}_{4} }

\begin{document}
   
\def\gC{\mbox{\boldmath $C$}}
\def\gZ{\mbox{\boldmath $Z$}}
\def\gR{\mbox{\boldmath $R$}}
\def\gN{\mbox{\boldmath $N$}}
\def\ua{\uparrow}
\def\da{\downarrow}
\def\a{\alpha}
\def\b{\beta}
\def\g{\gamma}
\def\G{\Gamma}
\def\d{\delta}
\def\D{\Delta}
\def\e{\epsilon}
\def\ve{\varepsilon}
\def\z{\zeta}
\def\h{\eta}
\def\th{\theta}
\def\k{\kappa}
\def\l{\lambda}
\def\L{\Lambda}
\def\m{\mu}
\def\n{\nu}
\def\x{\xi}
\def\X{\Xi}
\def\p{\pi}
\def\P{\Pi}
\def\r{\rho}
\def\s{\sigma}
\def\S{\Sigma}
\def\t{\tau}
\def\f{\phi}
\def\vf{\varphi}
\def\F{\Phi}
\def\c{\chi}
\def\w{\omega}
\def\W{\Omega}
\def\Q{\Psi}
\def\q{\psi}
\def\de{\partial}
\def\inf{\infty}
\def\ra{\rightarrow}
\def\bra{\langle}
\def\ket{\rangle}

\draft

%%%%%%%%%%%%%%%%%%
 \twocolumn[\hsize\textwidth\columnwidth\hsize\csname
 @twocolumnfalse\endcsname
%%%%%%%%%%%%%%%%%%

\widetext

\title{Electron-Phonon Interactions in the $W=0$ Pairing Scenario}

\author{Enrico Perfetto and Michele Cini}
\address{ Dipartimento di
Fisica,
Universita' di Roma Tor Vergata,\\ Via della Ricerca Scientifica,  1-00133 
Roma, Italy\\ and \\ INFN, Laboratori Nazionali di Frascati, C.P. 13, 00044 
Frascati, Italy  }

\maketitle

\begin{abstract}
We investigate the interplay of phonons and correlations in 
superconducting pairing by introducing  a  model Hamiltonian with on-site 
      repulsion  and  couplings to several vibration branches 
      having the Cu-O plane of the  cuprates as a 
      paradigm. We express the electron-phonon coupling (EP) through two 
      force constants for   O-Cu and O-O bond 
      stretchings.   Without 
      phonons, this reduces to the Hubbard Model, and      allows   purely electronic $W=0$
   pairing. A $W=0$ 
     pair is a two-body singlet eigenstate of the Hubbard Hamiltonian, 
  with no double occupancy, which gets bound from interactions with 
  background particles.   Indeed, this mechanism produces a  
      Kohn-Luttinger-like  pairing from the Hubbard repulsion,  
     provided  that its  symmetry is not severely distorted. 
    From the many-body theory, a canonical transformation  extracts 
      the effective two-body problem, which lends itself to 
      numerical analysis in case studies.  
      As a test, we use as a prototype system the $\cu$ cluster. 
          We show  analytically  that at weak EP coupling
      the additive contributions of the half-breathing modes reinforce 
      the electronic 
      pairing. 
      At intermediate and strong EP coupling and $U \sim t$, the model 
      behaves in a  
      complex and intriguing  way.  
\end{abstract}

\bigskip\bigskip\bigskip
%%%%%%%%%%%%%%%%%%
]
%%%%%%%%%%%%%%%%%%

\narrowtext

{\small 
%\tableofcontents

\section{Introduction}
\label{intro}

  While  the Fr\"ohlich mechanism of conventional superconductivity is driven by 
phonon exchange, the pairing mechanism in  highly correlated, 
narrow-band systems  could have a predominantly electronic origin\cite{dagotto}
and the Cu-O plane of cuprates is the most discussed 
example. Although this remains  a very controversial issue, most 
authors probably accept  at least the conceptual importance of a 
lattice counterpart of the Kohn-Luttinger  
idea\cite{kohn} that  attractive interactions result from mere repulsion. 
The Renormalization Group approach\cite{zanchietal} to the Hubbard 
Model shows that such a superconducting instability in the 
$d_{x^{2}-y^{2}}$ channel is  dominant near half filling, confirming the  results 
obtained with the FLEX approximation\cite{FLEX}.
One definition of  pairing is ${\tilde \Delta} < 0$, where  
\begin{equation}
{\tilde \Delta} (N+2) = E(N+2) + E(N)- 2E(N+1),
\label{tildedelta}
\end{equation}
and $E(N)$ is the ground state energy of the system with $N$  fermions;
this criterion is suitable for finite cluster calculations by exact 
diagonalization methods.  ${\tilde \Delta}<0$ was indeed observed in particular Hubbard 
clusters\cite{tosatti}\cite{white}, but  in many other examples with 
on-site repulsion on every site a large  ${\tilde \Delta} >0$
was found \cite{scalah} \cite{alascio}.  The $W=0$ 
theory\cite{PRB1997}\cite{PRB2002}\cite{PRB2003} gives a systematic 
method for producing and analizing examples of singlet pairing by repulsive 
interactions; it also allows  validating $|{\tilde \Delta}|$ as the pairing 
binding energy. In this framework   
the non abelian symmetry group of the underlying graph and the 
resulting degeneracy  is  crucial for the pairing mechanism. 
 The fillings and the symmetry channels where the $W=0$ pairing can occur are determined
in full generality by the $W=0$ theorem\cite{IJMPB2000}; these 
symmetries achieve  the same result as high  angular momentum and parallel spins in the 
Kohn-Luttinger\cite{kohn} continuum approach.

Anyhow, a purely electronic theory misses practically 
and conceptually important features of this complicated problem. 
First, many high-$T_{C}$ compounds exhibit a quite 
noticeable doping-dependent isotope effect\cite{crowford}, suggesting that 
electron-phonon (EP)  
interactions are important and should be included in the theory.
In addition, there is experimental evidence\cite{mcqueeney} that the
half breathing Cu-O bond stretching mode at $k=(\pi,0),(0,\pi)$ is significantly coupled 
with the doped holes in the superconducting regime and its 
contribution may be relevant for the $d_{x^{2}-y^{2}}$ 
pairing\cite{lanzara}\cite{gunnarsson}\cite{ishihara}.  A radical, 
yet  serious criticism of all electronic mechanisms was put forth by 
Mazumdar and coworkers\cite{mazumdar}. They suggested  that any  
pairing in Hubbard clusters is  of doubtful physical  interpretation due to the 
neglect  of the lattice degrees of freedom and the  Jahn-Teller (JT) effect. 
They argued that JT distorsions might well cause a larger  energy gain of the 
system with $N+1$ particles, and that could  reverse the sign of ${\tilde 
\Delta}$ obtained at fixed nuclei; in this case the pairing would be 
just an artifact of the Hubbard model.
This even led the authors to the conjecture that any ${\tilde \D}<0$ due to 
an electronic mechanism is just a finite size  effect, which vanishes for large 
systems like the JT effect does. Below, we shall show that the Mazumdar 
et al. argument\cite{mazumdar}, based on  a static Jahn-Teller approximation, can 
break down in a peculiar and nontrivial way when more flexible  wave functions 
are allowed; thus,  phonon pairing and $W=0$ pairing can be compatible, 
depending on the symmetries of both pair and vibration and on 
frequencies.

The Hubbard-Holstein  model, where electrons are coupled to a local Einstein phonon, is a simple way
to include both strong electronic correlations and  EP interactions.
Much is known about the possibility of a  superconducting phase
in this model.   Pao and Schuttler \cite{pao} applied the numerical renormalization group 
techniques within the FLEX approximation and found that in the square 
geometry $s$-wave pairing is enhanced by phonons, while 
$d_{x^{2}-y^{2}}$ pairing is suppressed. 
In the strong EP regime a Lang-Firsov\cite{lang} trasformation 
maps the Hubbard-Holstein model in an effective  Hamiltonian for 
hopping polarons with a screened on-site  interaction $\tilde{U}=U-{ 
g^{2}\over \omega}$,
where $U$ is the Hubbard repulsion, $\omega$ is the phonon frequency 
and $g$ is the EP coupling constant.  If overscreening is attained,  
$\tilde{U}<0$ becomes an effective attraction and  one gets bipolaronic bound 
states\cite{bipolaroni1}\cite{bipolaroni2}; however it is still unclear whether
they can exist as itinerant band states  \cite{alexandrov}\cite{bar}.
Petrov and Egami\cite{egami} found ${\tilde \Delta} <0$ in a 
doped 8-site Hubbard-Holstein ring at strong enough EP couling, 
while  otherwise the normal repulsion prevails. This situation is 
unavoidable in 1 dimensional repulsive Hubbard model, where no 
superconducting pairing exist.

In this paper we take the view that one of the sound   experimental 
facts  about  the CuO plane  in all cuprates is that the geometry  
permits $W=0$ pairs which  avoid completely the strong hole-hole 
repulsion. It is therefore highly plausible that such pairs are 
important ingredients of the theory, provided that there is a way out 
of the Mazumdar et al.\cite{mazumdar} argument.  Anyhow, it is not obvious that  the phonons
will reinforce the attraction while 
preserving the symmetry.  More generally, some vibrations could be pairing and 
others pair-breaking.
When lattice effects are introduced in the $W=0$  scenario,  
the situation is very different from the Petrov and Egami\cite{egami} model, when, as in
 the conventional (Fr\"ohlich) mechanism, phonons overscreen 
the electron repulsion; what happens if electronic screening already 
leads to pairing?   

To address these problems we use an 
extension of the Hubbard model in which bond stretchings dictate the 
couplings to the normal modes of the $C_{4v}$-symmetric configuration. This is 
physically more detailed than the  Hubbard-Holstein model, and does not 
restrict to on-site EP coupligs that would be impaired by a
strong Hubbard repulsion. 

The plan of the paper is the following. After introducing the model 
Hamiltonian in the next Section, we devote Sect.\ref{canonical} to a 
detailed derivation of  the effective interactions between holes in the $W=0$ pair, which is 
obtained by extending a previous Hubbard Model treatment. Our 
canonical transformation approach is  quite general for weak EP coupling and corresponds to 
the inclusion of all diagrams involving one-phonon and electron-hole 
pair exchange due to correlations. We specialize in Sect.\ref{clust} to 
the prototype  CuO$_{4}$ cluster,  describing 
electronic states and vibration modes.  The effective interaction is 
calculated explicitly in 
Sect.\ref{eff}. Next, we develop a theory 
based on the Jahn-Teller operator in Sect.\ref{JT?};  in this way we 
want to test the reliability of that approximation in modeling the 
behaviour of $W=0$ pairs in the presence of Jahn-Teller active modes.
The numerical results of the full 
theory are then exposed and discussed in Sect.\ref{num}; the exact 
data  for realistic vibration frequencies disagree from those of 
Sect.\ref{JT?} but are in  accord with the canonical 
transformation approach of Sect. \ref{eff}. The agreement is excellent 
at weak coupling, but the analytical approach is qualitatively 
validated also at intermediate coupling.   Finally 
Sect.\ref{Conclusions} is devoted to the conclusions.

\section{Model}\label{model}

We start from the  Hubbard model with on-site interaction $U$ and 
expand the hopping integrals $t_{i,j}({\bf r}_{i},{\bf r}_{j})$ in 
powers of the 
displacements ${\bf \r}_{i}$ around a $C_{4v}$-symmetric equilibrium configuration  
\begin{eqnarray}
t_{i,j}({\bf r}_{i},{\bf r}_{j}) \simeq t^{0}_{i,j}({\bf r}_{i},{\bf r}_{j})
+ \sum_{\a}\left[ \frac{\partial t_{ij}({\bf r}_{i},{\bf r}_{j}) }
{ \partial r^{\a}_{i}} \right] _{0}   \r^{\a}_{i}+\nonumber\\
+ \sum_{\a} \left[ \frac{\partial t_{ij}({\bf r}_{i},{\bf r}_{j}) }
{ \partial r^{\a}_{j}} \right] _{0}  \r^{\a}_{j}  
\; ,
\label{varhop}
\end{eqnarray}
where  $\a =x,y$. 
Below, we  write down the  $\r^{\a}_{i}$  in terms of the normal modes  
$q_{\eta \, \nu}$:
$
\r^{\a}_{i}=\sum_{\eta \, \nu} S^{\a}_{\eta \, \nu}(i) \; q_{\eta \, \nu} 
$, where $\eta$ is the label of   an irreducible representation ({\em irrep})  of the 
symmetry group of the undistorted system and $\n$ is a phonon branch.

Thus,  treating  the  Cu  atoms as fixed, for 
simplicity, one can justify an electron-lattice Hamiltonian:
\begin{equation}
H_{el-latt} =  H_{0} + V_{\rm tot} \, . 
\label{htot}
\end{equation}
Here $ H_{0} = H_{0}^{n}+H_{0}^{e}$ is given by
\begin{equation}
  H_{0} =\sum_{\eta} \hbar \omega_{\eta,\n} 
  b^{\dagger}_{\eta,\n} 
  b_{\eta,\n}+
  \sum_{i,j\s} t^{0}_{i,j}({\bf r}_{i},{\bf r}_{j})( c^{\dag}_{i \s}c_{j \s}+h.c),
  \end{equation}
where $\omega_{\eta,\n}$ are the  frequencies of the normal modes 
with creation operator $b^{\dagger}_{\eta,\n}$, while $c^{\dag}_{i 
\s}$ creates a fermion of spin $\s$ in site $i$.
Moreover, let $M$ denote  the O mass, $\xi_{\eta,\nu}=\lambda_{\eta  \, 
  \nu}\sqrt{\frac{\hbar}{2M\omega_{\eta,\n}}}$, with $\lambda_{\eta  \, \nu}$ 
  numbers of order unity 
that modulate the EP coupling strength.
Then, $ V_{\rm tot} = V+W$   reads
  \begin{eqnarray}
  V_{\rm tot}=\sum_{\eta,\n} 
  \xi_{\eta,\nu} ( 
b^{\dagger}_{\eta,\n}+ b_{\eta,\n}) H_{\eta, \n}+
U\sum_{i}n_{i\ua} n_{i\da},
\label{hel}
\end{eqnarray}

the  $H_{\eta,\n}$ operators  are given by
\begin{eqnarray}
 H_{\eta,\n}=\sum_{i,j} \sum_{\a, \sigma} \left\{ S^{\a}_{\eta \, \nu}(i)
 \left[ \frac{\partial t_{ij}({\bf r}_{i},{\bf r}_{j}) }
{ \partial r^{\a}_{i}} \right] _{0}  \right. \nonumber\\ \left.+ S^{\a}_{\eta \, \nu}(j)
\left[ \frac{\partial t_{ij}({\bf r}_{i},{\bf r}_{j}) }
{ \partial r^{\a}_{j}} \right] _{0}    \right\} ( c^{\dag}_{i, 
\s}c_{j ,\s}+h.c.) \, .
\label{heta}
\end{eqnarray}

In  previous work we have shown that the pure Hubbard model $H_{H}= H_{0}^{e}+W$
defined on the Cu-O plane and on the simple square as well
admits two-body singlet eigenstates with no
double occupancy on lattice sites. We refered to them 
as $W=0$ pairs. $W=0$ pairs are therefore  
eigenstates of the kinetic energy operator $H_{0}^{e}$  
and of the Hubbard repulsion $W$  with vanishing 
eigenvalue of the latter. The particles forming a $W=0$ pair have no 
direct interaction and are the main candidates to achieve bound 
states in purely repulsive Hubbard 
models\cite{EPJB1999}\cite{EPJB2000}\cite{EPJB2001}.
In order to study if the $W=0$ can actually form bound states in the 
many-body interacting problem, we developed a canonical 
transformation  of the Hubbard Hamiltonian\cite{EPJB1999}, which enables us to 
extract the effective interaction between the particles forming the 
pairs. Pairing was found in small symmetric clusters and large 
systems as well.

In the next Section we wish to derive  an 
effective interaction between the particles in the pair suitable for  $H_{el-latt}$,
by   generalizing 
the canonical transformation approach of Ref.\cite{EPJB1999}.

\section{Canonical Transformation}
\label{canonical}

 In this Section we assume that the system has periodic boundary conditions 
with  particle number  $N$; we  denote the phonon vacuum by  $|0\rangle \rangle$
and  the
non-interacting Fermi sphere by $|\Phi_{0}(N) \rangle $.  The creation operator of a $W=0$ pair is
obtained\cite{IJMPB2000} by 
applying an appropriate projection operator to 
$c^{\dagger}_{k\uparrow}c^{\dagger}_{k^{\prime}\downarrow}$, where 
the labels denote Bloch states. If we add a 
$W=0$ pair to
$|\Phi_{0}(N) \rangle \otimes |0\rangle \rangle$, the two extra particles, 
by definition, cannot 
interact directly (in first-order). 
Hence their effective interaction
comes out from virtual electron-hole (e-h) excitation and/or phonon exchange 
and in principle can be attractive.  
To expand 
 the interacting $(N+2)$-fermions 
ground state $|\Psi_{0}(N+2) \rangle$,
we build a complete set of configurations in the subspace
with vanishing  $z$ spin  component, considering the  vacuum 
state 
  $|\F_{0}(N)\rangle \otimes |0\rangle \rangle $ 
and  the   set   of excitations over it. 

We start by creating  $W=0$  pairs of  fermions over
$|\F_{0}(N)\rangle \otimes |0\rangle \rangle$; we 
denote with $|m\ket \otimes | 0 \rangle \rangle$ these states.
At weak coupling,   we  may  truncate the Hilbert space to the 
simplest excitations, i.e., to states involving 1 e-h pair or 1 
phonon created over  the $|m\ket \otimes | 0 \rangle \rangle$
states. 
We define the $|m\ket \otimes | q \rangle \rangle$ states,
obtained by creating a phonon denoted by $q=(\eta,\nu)$ over
the $|m\ket \otimes | 0 \rangle \rangle$ states. 
Finally we introduce
the $ |\alpha \rangle  \otimes | 0 \rangle \rangle$ states,
obtained from the $|m\ket \otimes | 0 \rangle \rangle$ states by creating 
1 electron-hole (e-h) pair. 

 The approximation can be systematically 
improved by including two or more electron-hole and excitations in 
the truncated Hilbert space, at the cost of heavier computation.

We now expand the interacting ground state in the truncated 
    Hilbert  space
\begin{eqnarray}
|\Psi _{0}(N+2)\ket={\sum_{m}}a_{m}|m\ket \otimes |0 \rangle \rangle 
+\nonumber\\ + {\sum_{m,q}}a_{m,q}|m\ket \otimes |q \rangle \rangle 
+{\sum_{\alpha }}
a_{\alpha }|\alpha \ket \otimes |0 \rangle \rangle  
\label{lungo}
\end{eqnarray}
and set up the Schr\"{o}dinger equation 
\begin{equation}
    H_{el-latt}  |\Psi _{0}(N+2)\ket = E \,|\Psi _{0}(N+2)\ket .  
    \label{sch}
\end{equation}
 
 We now
consider the effects of the operators $H_{0}$ and $V_{\rm tot}$ 
on the terms of $|\Psi _{0}(N+2)\ket$. Choosing the
$|m \rangle \otimes |0,q \rangle \rangle $,
$|\a \rangle \otimes |0 \rangle \rangle$ states to be eigenstates 
of the noninteracting term $H_{0}$ we have 
\begin{eqnarray}
H_{0} |m \rangle  \otimes |0 \rangle \rangle &=&
E_{m}  |m \rangle \otimes |0 \rangle \rangle,  \\
H_{0} |m  \rangle \otimes |q \rangle \rangle &=&
(E_{m} + \omega_{q} ) |m \rangle \otimes |q \rangle \rangle,\label{mq}  \\
H_{0} |\a  \rangle \otimes |0 \rangle \rangle&=&
E_{\a} |\a \rangle \otimes |0 \rangle \rangle \, .  
\label{Teffect}
\end{eqnarray}
Let us consider the action of $V$ and $W$ on the same states, 
taking in account that $V$ creates or annihilates up to 1 
phonon and 1 e-h pair, and $W$ is diagonal in the phonon states and can create or 
destroy up to 2 e-h pairs.
\begin{eqnarray}
(V+W) |m  \rangle \otimes |0 \rangle \rangle =
\sum_{m',q} V^{q}_{m,m'} |m'  \rangle \otimes |q \rangle 
\rangle + \nonumber \\
+ \sum_{m'}W_{m,m'} |m'  \rangle \otimes |0 \rangle 
\rangle +\sum_{\alpha}W_{m,\alpha} |\a  \rangle \otimes |0 \rangle \rangle  
\end{eqnarray}
\begin{eqnarray}
(V+W) |m  \rangle \otimes |q \rangle \rangle =
\sum_{m'} V^{q}_{m,m'} |m'  \rangle \otimes |0 \rangle \rangle + \nonumber \\
 +\sum_{\alpha}V^{q}_{m,\alpha} |\a  \rangle \otimes |0 \rangle \rangle 
 + \sum_{m'}W_{m,m'} |m'  \rangle \otimes |q \rangle \rangle
 \end{eqnarray}
\begin{eqnarray}
 (V+W) |\a  \rangle \otimes |0 \rangle \rangle =
\sum_{m',q} V^{q}_{\a,m'} |m'  \rangle \otimes |q \rangle 
\rangle +  \nonumber \\
+ \sum_{\a'}W_{\a,\a'} |\a'  \rangle \otimes |0 \rangle 
\rangle +\sum_{m}W_{\a,m} |m  \rangle \otimes |0 \rangle \rangle \, .
\label{Veffect}
\end{eqnarray}
The Schr\"{o}dinger equation yields three coupled equations for 
the coefficients $a$'s:
\begin{eqnarray}
(E_{m}-E)a_{m}+\sum_{m'}a_{m'}W_{m,m'}+\sum_{m',q}a_{m',q}V^{q}_{m,m'}+ \nonumber \\ 
+ \sum_{\a}a_{\a}W_{m,\a}=0 \, ;
\label{eqm}
\end{eqnarray}
\begin{eqnarray}
(E_{m}+\omega_{q}-E)a_{m,q}+\sum_{m'}a_{m',q}W_{m,m'}+\sum_{m'}a_{m'}V^{q}_{m,m'}+
\nonumber \\ + \sum_{\a}a_{\a}V^{q}
_{m,\a}=0 \, ;
\label{eqmq} 
\end{eqnarray}
\begin{eqnarray}
(E_{\a}-E)a_{\a}+\sum_{\a'}a_{\a'}W_{\a,\a'}+ \nonumber \\ + \sum_{m}a_{m} 
W_{\a,m}+\sum_{m,q}a_{m,q}V^{q}_{\a,m}=0 \, .
\label{eqa}
\end{eqnarray}
We define renormalized eigenenergies $E'_{\a}$  by taking a linear combination of 
the $\a$ states in such a way that
\begin{equation}
(H_{0}+W)_{\a,\a'}=
\d_{\a\a'}E'_{\a};
\label{alfapri}
\end{equation}
 Eq.(\ref{eqa}) can be solved for the $a_{\a}$ coefficients:
\begin{equation}
a_{\a}=\frac{1}{E'_{\a}-E}(\sum_{m}a_{m}W_{\a,m}+\sum_{m,q}a_{m,q}V^{q}_{\a,m}) \, .   
\label{adec}
\end{equation}    
Sobstituing $a_{\a}$ in Eq.(\ref{eqmq}) one gets:
\begin{eqnarray}
(E_{m}+\omega_{q}-E)a_{m,q} + 
\sum_{m'}a_{m',q}W_{m,m'} + 
\sum_{m'}a_{m'}V^{q}_{m,m'}+ \nonumber \\ 
+ \sum_{m',q'}a_{m',q'}(\sum_{\a} 
\frac{V^{q}_{m,\a}V^{q'}_{\a,m'}}{E'_{\a}-E} )+\quad\quad\nonumber\\
\sum_{m'}a_{m'}(\sum_{\a} \frac{V^{q}_{m,\a}W_{\a,m'}}{E'_{\a}-E} )
=0\quad\quad \, . 
\label{amq1}
\end{eqnarray}
Here, two important simplifications allow to proceed. First, as in 
Ref. \cite{PRB2002}, Eq.(49), one can show that
\begin{equation}
W_{m,m'}=W^{(d)}_{m,m'}+\d_{m,m'}W_{F},
\label{selfene}
\end{equation}
$W^{(d)}_{m,m'}$ is the direct interaction  among the 
particles forming the pair and  $W_{F}$ comes from the average
over the occupied states on the Fermi sphere and is a $m$-independent constant;
since $W^{(d)}_{m,m'}$ vanishes for the $W=0$ property, it holds
\begin{equation}
    \sum_{m'}a_{m',q}W_{m,m'} =  a_{m,q} W_{F}.
    \label{over}
\end{equation}
Moreover,
\begin{eqnarray}
\sum_{m',q'}a_{m',q'}(\sum_{\a} 
\frac{V^{q}_{m,\a}V^{q'}_{\a,m'}}{E'_{\a}-E} )= \nonumber \\
= a_{m,q} \sum_{\a} 
\frac{|V^{q}_{m,\a}|^{2}}{E'_{\a}-E} ) \, .
\label{self1}
\end{eqnarray}    
Indeed, $V$ is a one-body operator for the fermions; so  the 
electron-hole pair in the $\a$ state must be created by one $V$ 
factor and annihilated by the other; in this way, the $W=0$ pair is not 
touched. With these simplifications,  the 
contributions in Eqs.(\ref{over}) and (\ref{self1}) can be taken over 
to the l.h.s. of Eq.(\ref{amq1}), where they just renormalize   the 
eigen-energies 
of the $|m\rangle \otimes |q\rangle\rangle$ states. Thus, 
$E_{m}+\omega_{q} \rightarrow E'_{m}+\omega_{q}$, and 
the Eq.(\ref{amq1}) can easily solved for the $a_{m,q}$'s, as we did 
for the $a_{\a}$'s in Eq.(\ref{adec}):
\begin{equation}
a_{m,q}=\frac{1}{E'_{m}+\omega  
_{q}-E}\sum_{m'} a_{m'}(V^{q}_{m,m'}+\sum_{\a} 
\frac{V^{q}_{m,\a}W_{\a,m'}}{E'_{\a}-E}) \, .
\label{amq2}
\end{equation}  
Finally, substituting Eqs.(\ref{adec}, \ref{amq2}) into Eq.(\ref{eqm}), we can write
the Schr\"{o}dinger equation in terms of only the $|m \rangle$ states,
with the excitations-mediated interactions and with renormalized 
quantities: 
\begin{eqnarray}
0=(E_{m}-E)a_{m}  +a_{m} W_{F} +\quad\quad \nonumber\\  
+\sum_{m',m'',q}a_{m'}\frac{V^{q}_{m,m''}V^{q}_{m'',m'}}
{E'_{m''}+\omega  _{q}-E}+ \quad \quad \quad  \nonumber \\ 
+ 
\sum_{m',\a}a_{m'}\frac{W_{m,\a}W_{\a,m'}}{E'_{\a}-E}+\quad\quad\quad\quad 
\nonumber \\
+2\sum_{m',m'',q,\a} a_{m'} 
\frac{V^{q}_{m,m''}V^{q}_{m'',\a}W_{\a,m'}}{(E'_{m''}+\omega  
_{q}-E)(E'_{\a}-E)} \quad\quad\quad\nonumber \\ +
\sum_{m',m'',q,\a,\a'} a_{m'} \frac{W_{m, \a}V^{q}_{\a,m''}V^{q}_{m'',\a'}W_{\a',m'}}
{(E'_{m''}+\omega  _{q}-E)(E'_{\a}-E)(E'_{\a '}-E)} \, . \quad \quad
\label{coopereq}
\end{eqnarray}
Here, the last two terms are of higher order and must  be dropped; $E$ is the ground state of the system with $N+2$ Fermions; yet,
Eq.(\ref{coopereq}) is of the form of a Schr\"odinger equation with eigenvalue 
$E$ for the added pair.
We interpret $a_{m}$ as the expansion  coefficients over the $W=0$ pairs of the wave function of
the dressed pair, $|\varphi \rangle \equiv \sum_{m} a_{m} |m\rangle 
\otimes |0\rangle\rangle$.
This obeys the Cooper-like equation  
\begin{equation}
    H_{\rm pair}|\varphi \rangle  = 
E |\varphi \rangle  \label{coppieq}
\end{equation}
with the same $E$ as in Eq.(\ref{sch}), but with an effective two-body Hamiltonian:
\begin{equation}
H_{\rm pair} \equiv H_{0}+W_{F}+S[E] \, .
\label{sceq}
\end{equation}
Here $S$ is the $E-$dependent effective scattering operator
\begin{eqnarray}
S[E]_{m,m'}= \sum_{\a}
\frac{W_{m,\a}W_{\a,m'}}{E'_{\a}-E} +
\sum_{m'',q}\frac{V^{q}_{m,m''}V^{q}_{m'',m'}}
{E'_{m''}+\omega  _{q}-E} \, ,
\label{coopeff}
\end{eqnarray}
and therefore Eq.(\ref{coppieq}) must be solved self-consistently. 
Let us examine in detail the structure of the  $S[E]$ contribution. 
The matrix elements $S_{m,m'}$ may be written as
\begin{equation}
S_{m,m'}=(W_{\rm eff})_{m,m'}+F_{m}\d_{m,m'} \,.
\end{equation}
where $W_{\rm eff}$ is  the 
true effective interaction  between the 
electrons in the $m$ states, while the other term represents the forward scattering 
amplitude  $F$. 

The first-order self-energy $W_{F}\d_{m,m'}$ and the forward scattering 
term $F_{m}\d_{m,m'}$ are diagonal in the 
indices $m$ and $m'$, and therefore they renormalize the non-interacting 
energy $E_{m}$ of the $m$ states: 
\begin{equation}
    E_{m}\ra E^{(R)}_{m}=E_{m}+W_{F}+F_{m}.
\label{eneren}
\end{equation}
If the effective  interaction $W_{\rm eff}$  is attractive and produces 
bound, localized  states the 
spectrum of the Schr\"odinger equation with the Hamiltonian in Eq.(\ref{sceq}) contains 
discrete states below the unpaired states. In an extended system, we 
have  bound states below the threshold of the {\em continuum}.
The threshold may be defined (in clusters and in extended systems)  by 
\begin{equation}
E^{(R)}_{T}\equiv \min_{\{m\}}[ E^{(R)}_{m}(E)],
\label{min}
\end{equation}
which, according to Eq.(\ref{eneren}), takes into account all the pairwise interactions except those between 
the particles in the pair.  Note that this is an extensive quantity, 
i.e. an $N+2$-particle energy.
The ground state energy $E$ may be conveniently written as 
\begin{equation}
   E=  E^{(R)}_{T}+\D ;
    \label{notilde}
\end{equation}
 $\D<0$ indicates a Cooper-like instability of the normal Fermi liquid 
and its magnitude  represents the binding energy of the pair. 
 
Below, we solve Eq.(\ref{coppieq}) explicitly for the CuO$_{4}$ 
cluster with open boundary conditions, where the above theory is 
readily applied.

\section{Prototype Cluster}
\label{clust}

As an illustrative application of the above pairing scheme, in this preliminary 
work  we focus  on CuO$_{4}$, the smallest cluster yielding $W=0$ pairing 
in the Hubbard model. This requires 4 holes, (total number, not 
referred to half filling); such a doping is somewhat unrealistic, but  larger 
$C_{4v}$-symmetric clusters and the full CuO$_{2}$ plane also show $W=0$ pairing in 
the doping regime relevant for cuprates\cite{PRB1997}\cite{EPJB1999}. 
Remarkably, in the pure Hubbard model, one can verify that $\D=\tilde{\D}(4)$ at least at 
weak coupling\cite{PRB1997}, which demonstrates that $\tilde{\D}$ has 
the physical meaning of an effective interaction. 
CuO$_{4}$  represents a  good test of the interplay 
between electronic and phononic pairing 
mechanisms since we can compare exact diagonalization 
results with the analytic approximations of the canonical 
transformation. A further merit of this model is that it demonstrates 
dramatically the decisive role of symmetry in the electronic pairing 
mechanism: any serious distortion of the square symmerty restores the 
normal $\tilde{\D}>0$ situation\cite{PRB1997}. Since vibrations cause 
distortions it is not evident {\em a priori} that they tend to help 
pairing;  in particular we may expect that Jahn-Teller distortions are 
going to prevent $W=0$ pairing altogether. On the other hand, the 
Fr\"ohlich mechanism of conventional superconductivity is based on 
phonon exchange. This suggests that the role of EP coupling  is 
complex.

CuO$_{4}$ allows   only the coupling to phonons at the centre 
or at  the edge of 
the Brillouin Zone;  however, phonons near the edge are precisely those  most 
involved\cite{mcqueeney}\cite{lanzara}. 
Even in this small system the virtually 
exact diagonalizations are already hard and the next 
$C_{4v}$-symmetric example, the 
Cu$_{5}$O$_{4}$ cluster\cite{EPJB2000},
is  much more demanding for the number of vibrations and the size 
of the electronic Hilbert space. 
\begin{figure}[H]
\begin{center}
       \epsfig{figure=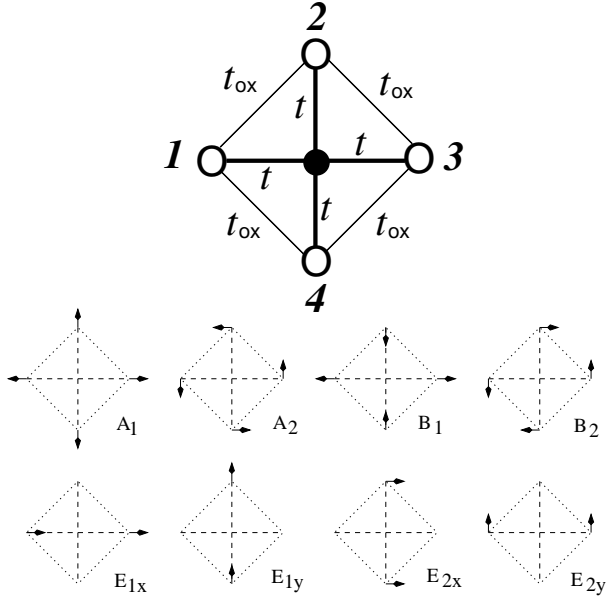,width=8cm}
       \caption{\footnotesize{
       Pictorial representation of the ionic displacements in the eight normal modes of the CuO${4}$ 
       cluster, labelled according to the {\em irreps} of the $C_{4v}$ Group. }}
 \label{cuo4}
\end{center} 
\end{figure}

Starting with the $C_{4v}$-symmetric arrangement, 
any   displacement of the Oxygens in the plane can be analised in 
{\em irreps}, 
$A_{1},A_{2}, B_{1}, B_{2}, E_{1},E_{2}$, see Fig.\ref{cuo4}.

We suppose that the hopping integrals depend only on the bond 
lengths\footnote{
Some Authors use an alternating sign convention for the bonds from a 
given Cu site. However, this is just a gauge; in the present 
CuO$_{4}$ case, this corresponds to changing the sign of two 
opposite Oxygen orbitals. Even in the full plane, starting from  
positive $t$ integrals, one can  introduce staggered signs by  
 negating a sublattice of O orbitals; then, one can arrange opposite 
 signs for the bonds of each O by simply negating a sublattice of Cu.
All this has no physical implications, and in our opinion does not help to visualize 
the real symmetry of the problem.
}. 
Hence, the EP coupling is expressed through just two parameters $g$ and  
$ g_{ox}$, defined as follows:
denoting e.g. by $t^{1}$ the hopping integral between Oxygen 1 and the Cu 
and by $t_{ox}^{1,2}$ the one between Oxygens 1 and 2 (see Fig.\ref{cuo4}),
$
g \equiv [\frac{\partial t^{1}}{\partial {|\bf r}_{1}| 
}]_{0}
$ and  
$
 g_{ox} \equiv [\frac{\partial t_{ox}^{1,2}}{\partial {|\bf r}_{1}- {\bf r}_{2}|}]_{0}.
$ 
We take $ g <0$ since a positive Cu-O hopping integral
decreases as the Cu-O distance is increased. On the other hand
$ g_{ox} >0$, since physically the O-O hopping integral
has the opposite sign with respect the Cu-O one.
Following 
Eqs.(\ref{htot}-\ref{heta}), the second-quantized electron-lattice Hamiltonian reads
\begin{eqnarray}
&&H_{el-latt}^{{\rm CuO_{4}}} 
=\varepsilon_{p}\sum_{i,\s} n_{i,\s}+\varepsilon_{d}\sum_{\s}n_{d,\s}+\nonumber\\
&&\sum_{\eta} \hbar \omega_{\eta} 
  b^{\dagger}_{\eta} 
  b_{\eta}+t\sum_{i\s}( d^{\dag}_{\s}p_{i\s}+h.c)\nonumber \\
&+& t_{ox}\sum_{i\,\s}( p^{\dag}_{i\s}p_{i+1\s}+ h.c.)+ 
U(\sum_{i}n^{(p)}_{i\ua}\hat{n}^{(p)}_{i\da}+
\hat{n}^{(d)}_{\ua}\hat{n}^{(d)}_{\da}) + \nonumber \\ &+& \sum_{\eta} 
 \x_{\eta}  ( 
b^{\dagger}_{\eta}+ b_{\eta}) H_{\eta}; 
\label{hsequ}
\end{eqnarray}
where $p^{\dag}_{i\s}$ and $p_{i\s}$ are the hole creation and 
annihilation operators onto the oxygen $i=1,..,4$ with spin 
$\s=\ua,\da$, $d^{\dag}_{\s}$ and $d_{\s}$ are the hole creation and 
annihilation operators onto the central copper site,  
while $n^{(p)}_{i\s}=p^{\dag}_{i\s}p_{i\s}$ and 
$n^{(d)}_{\s}=d^{\dag}_{\s}d_{\s}$ are the 
corresponding number operator. Henceforth we set
$\varepsilon_{p}=\varepsilon_{d} = 0$ for 
convenience, since this simple choice is adequate for the present 
qualitative purposes. 
Also, we are assuming for simplicity that the Oxygen-Oxygen hopping $t_{ox}$
is zero, and O-O hoppings are important only once the ions are moved.  Similar 
results are obtained using  a realistic $t_{ox}$, except that pair binding 
energies are somewhat reduced.

The $H_{\eta}$ matrices are given by
\begin{eqnarray}H_{A_{1}}&=& \frac{1}{2}  g \,\sum_{i\s}  
(d^{\dag}_{\s}p_{i\s}+h.c.)+ \nonumber\\
&&\frac{1}{\sqrt{2}}   g_{ox} \, \sum_{i\,\s} 
(p^{\dag}_{i\s}p_{i+1\s}+ h.c.) \, ;  \nonumber \\
H_{A_{2}}&=& 0 \, ;  \nonumber \\
H_{B_{1}}&=& \frac{1}{2} g 
\,\sum_{\s}(d^{\dag}_{\s}p_{1\s}-d^{\dag}_{\s}p_{2\s}+d^{\dag}_{\s}p_{3\s}-d^{\dag}_{\s}p_{4\s}+h.c.) \, ; \nonumber \\
H_{B_{2}}&=& \frac{1}{\sqrt{2}}  g_{ox} \, \sum_{\s} (
p^{\dag}_{1\s}p_{2\s}-p^{\dag}_{2\s}p_{3\s}+ \nonumber\\ 
&&p^{\dag}_{3\s}p_{4\s}-p^{\dag}_{4\s}p_{1\s}+ h.c.) \, ; \nonumber \\
H_{E_{1x}}&=& \frac{1}{\sqrt{2}} g 
\,\sum_{\s}(-d^{\dag}_{\s}p_{1\s}d^{\dag}_{\s}p_{3\s}+h.c.)+  \nonumber \\ 
 \frac{1}{2}  g_{ox} \, \sum_{\s}&& (
-p^{\dag}_{1\s}p_{2\s}+p^{\dag}_{2\s}p_{3\s}+p^{\dag}_{3\s}p_{4\s}-p^{\dag}_{4\s}p_{1\s}+ h.c.) \, ; \nonumber \\
H_{E_{1y}}&=& \frac{1}{\sqrt{2}} g 
\,\sum_{\s}(d^{\dag}_{\s}p_{2\s}-d^{\dag}_{\s}p_{4\s}+h.c.) +  \nonumber \\
 \frac{1}{2}  g_{ox} \, \sum_{\s} &&(
p^{\dag}_{1\s}p_{2\s}+p^{\dag}_{2\s}p_{3\s}-p^{\dag}_{3\s}p_{4\s}-p^{\dag}_{4\s}p_{1\s}+ h.c.) \, ; \nonumber\\
H_{E_{2x}}&=&  \frac{1}{2}  g_{ox} \, \sum_{\s} (
p^{\dag}_{1\s}p_{2\s}-p^{\dag}_{2\s}p_{3\s}-p^{\dag}_{3\s}p_{4\s} \nonumber \\
&&+p^{\dag}_{4\s}p_{1\s}+ h.c.) \, ; \nonumber \\
H_{E_{2y}}&=& \frac{1}{2}  g_{ox} \, \sum_{\s} (
-p^{\dag}_{1\s}p_{2\s}-p^{\dag}_{2\s}p_{3\s} \nonumber  \\ 
&&+p^{\dag}_{3\s}p_{4\s}+p^{\dag}_{4\s}p_{1\s}+ h.c.) \, .  
\label{modnor}
\end{eqnarray}

In order to make contact with the physics of cuprates,  let us discuss
the connection between the normal modes of the CuO$_{4}$ cluster andthe phonon modes of the Cu-O planes.
There is experimental evidence\cite{lanzara} that the possibly relevant modes for 
superconductivity lie on the CuO$_{2}$ planes and have a
Cu-O bond stretching origin. In particular the LO {\it half-breathing } 
mode with $k=(\pi,0),(0,\pi)$ is believed to couple significantly 
with the doped holes in the superconducting  regime.
In the CuO$_{4}$ cluster the half breathing modes are contained in 
the breathing mode $A_{1}$ and in the quadrupolar mode $B_{1}$ by 
means of the linear combination $q_{A_{1}}\pm q_{B_{2}}$. 
We argue that qualitatively  the effect of the  coupling with the $A_{1}$ and $B_{1}$ modes 
 should give us 
clues about the interplay between  electronic 
$W=0$ pairing and phonon exchange.

\section{Lowest-order Effective interaction in CuO$_{4}$}\label{eff}

The mere Hubbard CuO$_{4}$ cluster with O-O 
hopping    $t_{ox}=0$ yields\cite{PRB1997} ${\tilde \Delta} (4) <0$, 
due to a couple of degenerate   $W=0$ bound pairs,  in the $A_{1}$   
and  $B_{2}$ {\em irreps} of the $C_{4v}$ group; therefore in Eq.(\ref{coopeff})
we set the $m=m'$  labels accordingly. 
At weak coupling, we may simplify  Eq.(\ref{coopeff}), neglecting all 
renormalizations;  the phonon-mediated 
interaction  for the   $B_{2}$ pair reads:
\begin{equation}
\sum_{m'',q}\frac{V^{q}_{B_{2},m''}V^{q}_{m'',B_{2}}}
{E'_{m''}+\omega ' _{q}-E} 
=-4 g_{ox}^{2}
\frac{\lambda_{B_{2}}^{2}}{2\varepsilon_{A_{1}}+\omega_{B_{2}}-E} \;.
\label{phonb2}
\end{equation}
Note that in the denominator in the r.h.s., 
 $2\varepsilon_{A_{1}}+\omega_{B_{2}}$ is the energy of an unrenormalized excited 
$|m\rangle\otimes | q\rangle\rangle$  state, which at weak 
couplig is higher than the ground state energy $E$; hence the r.h.s. 
must be negative and the $B_{2}$ phonon is synergic with electronic 
pairing. On the other hand, the vibronic effective interaction for the $A_{1}$ pair is:
\begin{eqnarray}
\sum_{m'',q}\frac{V^{q}_{A_{1},m''}V^{q}_{m'',A_{1}}}
{E_{m''}+\omega ' _{q}-E} 
=  -\frac{4}{3} g_{ox}^{2}
\left( \frac{\lambda_{B_{2}}^{2}}{2\varepsilon_{A_{1}}+\omega_{B_{2}}-E} +           
\right. \nonumber\\ \left. 
\frac{2 \lambda_{A_{1}}^{2}}{2\varepsilon_{A_{1}}+\omega_{A_{1}}-E}
-\frac{\lambda_{E_{1}}^{2}}{2\varepsilon_{A_{1}}
+\omega_{E_{1}}-E}-\frac{\lambda_{E_{2}}^{2}}{2\varepsilon_{A_{1}}+\omega_{E_{2}}-E}\right). 
 \label{phona1}\end{eqnarray}
This  shows that in the $A_{1}$ sector
the total sign depends on the relative weight of attractive and 
repulsive contributions. 
Eqs.(\ref{phonb2},\ref{phona1}) show that at weak coupling  
$A_{1}$ and $B_{2}$ modes 
are synergic to the $W=0$ pairing, while both longitudinal and 
transverse $E$ modes are 
pair-breaking.  The half-breathing  modes that are deemed most 
important\cite{lanzara}\cite{gunnarsson} are $A_{1}\pm B_{1}$ 
combinations, but $B_{1}$ does not appear in 
Eqs.(\ref{phonb2},\ref{phona1}). The numerical calculations reported below 
confirm these findings in a broad range of parameters.

For the sake of argument, in the explicit calculations we took all the normal modes 
with the same  energy 
$\varepsilon_{0}=\hbar \omega_{0} = 10^{-1} \rm{eV}$ and 
$\lambda_{\eta} = 1$. 
This  sets the  length scale of lattice effects
$\xi_{0}=\sqrt{\frac{\hbar}{2M \omega_{0}}}\simeq 10^{-1} \stackrel{\circ}{\rm{A}} 
$
where we used $M=2.7 \times 10^{-26}$Kg for Oxygen.

With this choice,  the Cooper-like equation (\ref{coopereq}) reads
\begin{eqnarray}
(2\varepsilon_{A_{1}}-E)&-&\frac{U^{2}}{16}
\left(\frac{1}{\varepsilon_{B_{1}}+\varepsilon_{A_{1}}-E} -
\frac{1}{2}\frac{1}{\varepsilon_{A_{1}}+\varepsilon_{A'_{1}}-E}\right)- 
\nonumber \\
&-&\frac{4}{3} g_{ox}^{2} \,
\frac{1}{2\varepsilon_{A_{1}}+\omega_{0}-E}=0 
\label{coopfina1}
\end{eqnarray}
in the $ A_{1}$ channel and
\begin{eqnarray}
(2\varepsilon_{A_{1}}-E)&-&\frac{U^{2}}{16}
\left(\frac{1}{\varepsilon_{B_{1}}+\varepsilon_{A_{1}}-E} -
\frac{1}{2}\frac{1}{\varepsilon_{A_{1}}+\varepsilon_{A'_{1}}-E}\right) - 
\nonumber \\
&-&4 g_{ox}^{2} \,
\frac{1}{2\varepsilon_{A_{1}}+\omega_{0}-E}=0  \, .
\label{coopfinb2}
\end{eqnarray} 
for $B_{2}$ pairs. The eigenvalue $E$, like in Eq.(\ref{sch}), 
is the total energy of the cluster; it must be compared with the threshold 
$E^{(R)}_{T}$ of Eq.(\ref{min}), whose noniteracting limit is $E^{(R)}_{T} 
= 2\varepsilon_{A_{1}}$  since the  
degenerate level energy (see Table II,  Appendix A) is
$\varepsilon_{p}=0$.
It turns out that using Eq.(\ref{notilde}) in  the weak coupling 
approximation, that ignores renormalizations,  the effective interaction  is 
$\Delta=E-2\varepsilon_{A_{1}}$; in Appendix B we verify by perturbation theory
that like in the Hubbard model, $\D=\tilde{\D}(4)$; this supports our 
interpretation of $\tilde{\D}$ as minus the pairing energy.

The trend of   $\Delta$ in both channels  is shown in Fig.\ref{bind}.
The vibrations split the degeneracy of the $W=0$ pairs,
effectively lowering the symmetry like a nonvanishing  
$t_{ox}$.
Pairing is enhanced in the $A_{1}$ sector as well, albeit less than in $B_{2}$; 
without phonons,  $\D \simeq -20$meV for both the $W=0$ 
pairs.

\begin{figure}[H]
\begin{center}
	\epsfig{figure=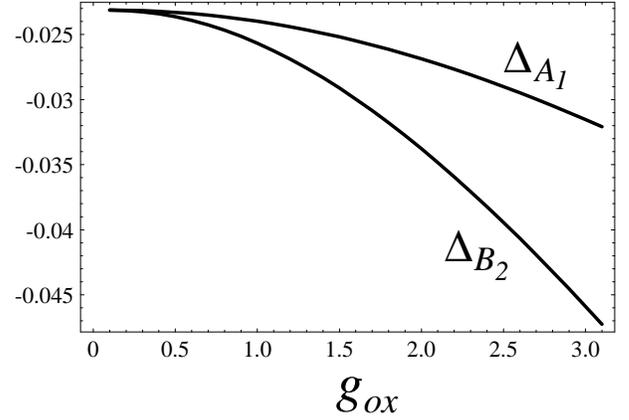,width=8cm}
	\caption{ \footnotesize{ Analytical results of the canonical transformation: 
	pair binding energy in the $A_{1}$ and $B_{2}$ sectors as a 
	function of $ g_{ox} $.
	Here we used $\lambda_{\eta}=1$ for every mode, $t=1$eV, $t_{ox}=0$, $U=1$eV;
	$ g_{ox} $ is in  units of  
	$\varepsilon_{0}/\xi_{0}=1$eV$\times 
	\stackrel{\circ}{\rm{A}}^{-1} $, $\D$ is in eV.}}
  \label{bind}
\end{center} 
\end{figure}

 \section{Jahn-Teller mixing of electronic ground states and pairing}
\label{JT?}
If really  the inclusion of the
lattice degrees of freedom systematically leads to ${\tilde \D}>0$, 
purely electronic cluster models become totally irrelevant to 
superconductivity, as it was argued \cite{mazumdar}. 
However small, the CuO$_{4}$ cluster yields electronic pairing and 
allows to  test this important point. 

In this Section we set up a conventional calculation of the JT effect 
involving degenerate electronic ground states and their mixing with the 
vibrations. 
We first take the nuclei as  frozen in a $C_{4v}$-symmetric configuration  
and diagonalize the purely electronic  part of the Hamiltonian: 
\begin{eqnarray}
H_{el}^{{\rm CuO_{4}}}=t\sum_{i\s}( 
d^{\dag}_{\s}p_{i\s}+h.c)+\nonumber\\
U(\sum_{i}n^{(p)}_{i\ua}n^{(p)}_{i\da}+
n^{(d)}_{\ua}n^{(d)}_{\da}) \, .
\label{senzatox}
\end{eqnarray}
As before, we are using  $t_{ox}=\varepsilon_{p}=\varepsilon_{d} =0 $.

  The JT effect  arises if the ground 
state of $H_{el}^{{\rm CuO_{4}}}$ is degenerate, with  a  ground state multiplet
$\left\{|\Psi_{1}\rangle , \ldots, |\Psi_{n} \rangle \right\}$ such 
that $H_{el}^{{\rm CuO_{4}}} |\Psi_{k}\rangle = E_{0} |\Psi_{k}\rangle $. If we 
take matrix elements of  $H_{el-latt}^{{\rm CuO_{4}}}$ in this 
truncated ($n$-dimensional) electronic basis integrating over electrons and keeping 
boson operators we get the  {\it dynamical} JT 
Hamiltonian,\cite{bersuker},\cite{grosso} with matrix 
elements:
\begin{equation}
H^{JT}_{\alpha,\beta}=( E_{0} + \sum_{\eta} \hbar \omega_{\eta} b^{\dagger}_{\eta} b_{\eta})\,
\delta_{\alpha,\beta} + 
\langle \Psi_{\alpha} |V|\Psi_{\beta} \rangle \, .
\label{jtpb}
\end{equation}
It is worth noting  that neglecting  the nuclear kinetic energy 
(i.e. $-\frac{\hbar^{2}}{2M} \sum_{i} \frac{\partial ^{2}}{\partial^{2} 
{\bf r}_{i}} \rightarrow 0$  ) and treating the nuclear positions as 
variational parameters corresponds to  the   {\it static} JT 
Hamiltonian, but we follow the dynamic treatment which is superior.

In the following we assume that the initial configuration is stable with
respect to the mode $A_{1}$ which
only changes the scale of the CuO$_{4}$ molecule. Since this mode  
does not produce any JT distorsion, it is not involved in the 
arguments of Ref.\cite{mazumdar}. In this Section, we study   
$\tilde{\D}(4)$ in this approximation, according to Eq.(\ref{tildedelta}).
The ground state with 2 holes is a nondegenerate totalsymmetric
singlet unaffected by  the JT effect; in the other cases, the use of 
the Hamiltonian (\ref{jtpb}) is justified provided that the excited 
states are several phonon energies above the ground state.

\subsection{Three-hole ground state mixing}

With three holes the ground state belongs to the 3-dimensional irrep    of 
$S_{4}$ which  in C$_{4v}$ breaks into
$B_{1} \oplus E$. To illustrate the electronic structure and its 
dependence on distortions, in Fig.\ref{apesb2} we show the adiabatic potential 
energy surface  projected along the $B_{2}$ distortion. 
 Projecting on the other directions, we obtain similar trends. It is clear that the ground 
state multiplet is well separated from the excited states and hence 
this treatment of the  JT effect is well justified.

\begin{figure}[H]
\begin{center}
	\epsfig{figure=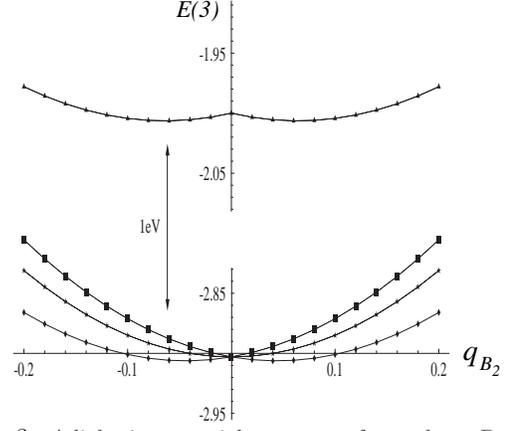,width=6.5cm}
	\caption{\footnotesize{Adiabatic potential energy surfaces along $B_{2}$ for
	the ground and first excited   3-hole states of $CuO_{4}$. Here, 
	$q_{B_{2}}$ denotes the classical normal coordinate; 
	$U=5$eV, $t=1$eV, $ g =-2.4$, $ g_{ox}=0.6$ in units 
	eV/$ \stackrel{\circ}{\rm{A}}$ and 
	$\omega_{\eta}=0.1$eV $\forall \, \eta$; the dispacement $q_{B_{2}}$ 
	is in $ \stackrel{\circ}{\rm{A}}$, energies are in eV.
	The ground state multiplet is below the first excited state by $\sim 1$eV}.}
    \label{apesb2}
\end{center} 
\end{figure}

Since $E \otimes E$  contains all the irreps of  $C_{4v}$, all the normal modes are JT active
in this  case.
Following Eqs.(\ref{jtpb}), we computed the following $V$ matrix elements 
in the 3-hole  ground state multiplet with  the Hubbard interaction, 
using Eq.(\ref{modnor}). The 4 independent elements at 
the optimal value $U/t \sim 5$, where the $W=0$ pair binding energy is
maximum, are:
\begin{eqnarray}
\gamma_{1}&=&  \langle \Psi_{B_{1}} | H_{E_{1x}} |\Psi_{E_{x}} \rangle = 
0.17 \,  g_{ox}\, \label{first} \\ 
\gamma_{2}&=&  \langle \Psi_{B_{1}} | H_{E_{2y}} |\Psi_{E_{x}} \rangle = 
0.24 \,  g + 0.17 \,  g_{ox} \\ 
\gamma_{3}&=&  \langle \Psi_{E_{x}} | H_{B_{1}} |\Psi_{E_{x}} \rangle = 
0.24 \,  g  \,\label{comecipare} \\ 
\gamma_{4}&=& \langle \Psi_{E_{y}} | H_{B_{2}} |\Psi_{E_{x}} \rangle = 
-1.05 \,  g_{ox} \label{last}
\end{eqnarray}
The JT Hamiltonian reads 
\begin{eqnarray}
    H_{JT}(3)=  \left[  E_{0}(3) +
    \sum_{\eta} \hbar \omega_{\eta} \, b^{\dagger}_{\eta} b_{\eta}
    \right] \otimes {\bf 1}_{3\times3} +\nonumber\\
    + \sum_{\eta} \x_{\eta}(b^{\dagger}_{\eta}
    +b_{\eta}) M_{\eta}
    \end{eqnarray}
where $E_{0}(3)$ is the ground state energy of $H_{el}^{{\rm CuO_{4}}}$  
with 3 holes and
\begin{equation}
    M_{B_{1}} =\gamma_{3}\left(
\begin{array}{ccc}
0&0&0\\
0&  1&0\\
0&0&-1
\end{array}
\right),
\end{equation}
\begin{equation}
    M_{B_{2}} =\gamma_{4} \left(
\begin{array}{ccc}
0&0&0\\
0& 0& 1\\
0& 1&0
\end{array}
\right) 
\end{equation}
\begin{equation}
    M_{E_{1x}} = \left(
 \begin{array}{ccc}
 0&\gamma_{2}&-\gamma_{1}\\
 \gamma_{2}& 0& 0\\
 -\gamma_{1}& 0&0
 \end{array}
 \right),
 \end{equation}
 \begin{equation}
     M_{E_{1y}} =  \left(
 \begin{array}{ccc}
 0&-\gamma_{1}&-\gamma_{2}\\
 -\gamma_{1}& 
 0& 0\\
 -\gamma_{2}& 0&0
 \end{array}
 \right)  
 \end{equation}
 \begin{equation}
     M_{E_{2x}} = \gamma_{1} \left(
 \begin{array}{ccc}
 0&-1 &1\\
 -1 & 
 0& 0\\
 1 & 0&0
 \end{array}
 \right),
 \end{equation}
 and
 \begin{equation}
     M_{E_{2y}} =\gamma_{1} \left(
 \begin{array}{ccc}
 0& 1 & 1 \\
 1 & 
 0& 0\\
 1 & 0&0
 \end{array}
 \right) .
 \end{equation}

We numerically diagonalized 
$H_{JT}(3)$  in the Hilbert space spanned by $| \Psi_{\eta} \rangle 
\otimes | \Phi_{N_{ph}} \rangle$, where $\eta=B_{1},E_{x},E_{y}$ is the 
electronic state and $ | \Phi_{N_{ph}} \rangle$
is a vibration state in the truncated Hilbert space with all modes 
having  vibrational 
quantum numbers $\leq N_{ph}$. Excluding the breathing mode the size of the 
problem is $3 (N_{ph}+1)^{6}$. 
We consider $N_{ph} = 3$, since  already in the weak coupling regime  $\tilde{\Delta}$ changes sign.
We studied $\omega_{\eta} =\omega = 0.1 eV,\; \lambda_{\eta} = 1\; \forall \eta$ in the range of  $| g |$ and $| g_{ox}|$  
between   $0$ and $2.4$eV/$\stackrel{\circ}{\rm{A}}$, which means that
the  number ratios $|\gamma_{i}| \sqrt{\frac{\hbar}{2M\omega}}/ \hbar \omega$
vary between $0$ and $0.5$.   This condition ensures that the weak coupling 
regime behooves and $N_{ph}=3$ is indeed adequate. The results are shown in 
Fig.\ref{ener3h}.  
\begin{figure}[H]
\begin{center}
	\epsfig{figure=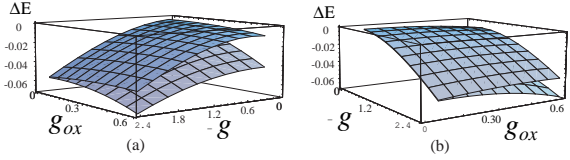,width=7.5cm}
	\caption{\footnotesize{(a) Vibronic correction $\D E$ to $E_{0}(3)$ 
	to the $E$ state (lower surface) and $B_{1}$ state (upper one)
	as a function of $ g $ and $ g_{ox}$. (b)  the same 
	surfaces  
	from another point of view.
	Here, $t=1$eV, $U=5$eV,  $\Delta E$ is eV of $\hbar \omega$, $ g $
	and $ g_{ox}$ are in units of 
	$\varepsilon_{0}/\xi_{0}=1$eV$\times \stackrel{\circ}{\rm{A}}^{-1} $.}}
        \label{ener3h}
\end{center} 
\end{figure}

The lower surface represents the ground state energy shift $\Delta E$ 
 for electronic states belonging to   the degenerate irrep $E$,
 while the higher one is $\Delta E$ refers to the $B_{1}$ state.
The JT effect partially removes the three-fold degeneracy
and the ground state is a $E$ doublet. Note that according to the 
textbook, static  JT effect, one should observe a total removal of 
the degeneracy. This is however not borne out by the dynamical 
calculation and,   for $ g_{ox}=0$, all 
the three states remain degenerate (see Fig.\ref{ener3h}.b).

The way the system dynamically distorts is also of interest. 
The only  vibration having the coordinate on
the diagonal of  
$H_{JT}(3)$ is  $B_{1}$; thus,  the $E$ doublet can only distort along 
the $B_{1}$ normal mode. 
In other terms, with $\hat{q}_{B_{1}} = \x_{B_{1}} 
(b_{B_{1}}^{\dagger}+b_{B_{1}})$,   $\langle \hat{q}_{B_{1}} \rangle \equiv \langle 
\Psi^{0}_{E_{x}} | \hat{q}_{B_{1}} |
\Psi^{0}_{E_{x}} \rangle =
- \langle \Psi^{0}_{E_{y}} | \hat{q}_{B_{1}} |  \Psi^{0}_{E_{y}} \rangle 
\neq 0$. 
The trend of $\langle \hat{q}_{B_{1}} \rangle$ as a function of $ g $ 
and $ g_{ox}$ is shown in  Fig.\ref{spostb1}. We observe that
$\langle \hat{q}_{B_{1}} \rangle \to 0$ as $ g \to 0$; we also 
remark that the deformation depends essentially by $ g $ and
only weakly  on $ g_{ox}$ because according to Eq.(\ref{comecipare}) 
the coupling constant  $\gamma_{3}$ responsible for the distorsion along
$B_{1}$ depends on 
$ g $ and not on $ g_{ox}$.  The 
$E$-$B_{1}$ splitting, on the contrary, depends on $ g_{ox}$ 
and only weakly on $ g $. The naive expectation that splittings 
go along with distortions only holds for static ones.
\begin{figure}[H]
\begin{center}
	\epsfig{figure=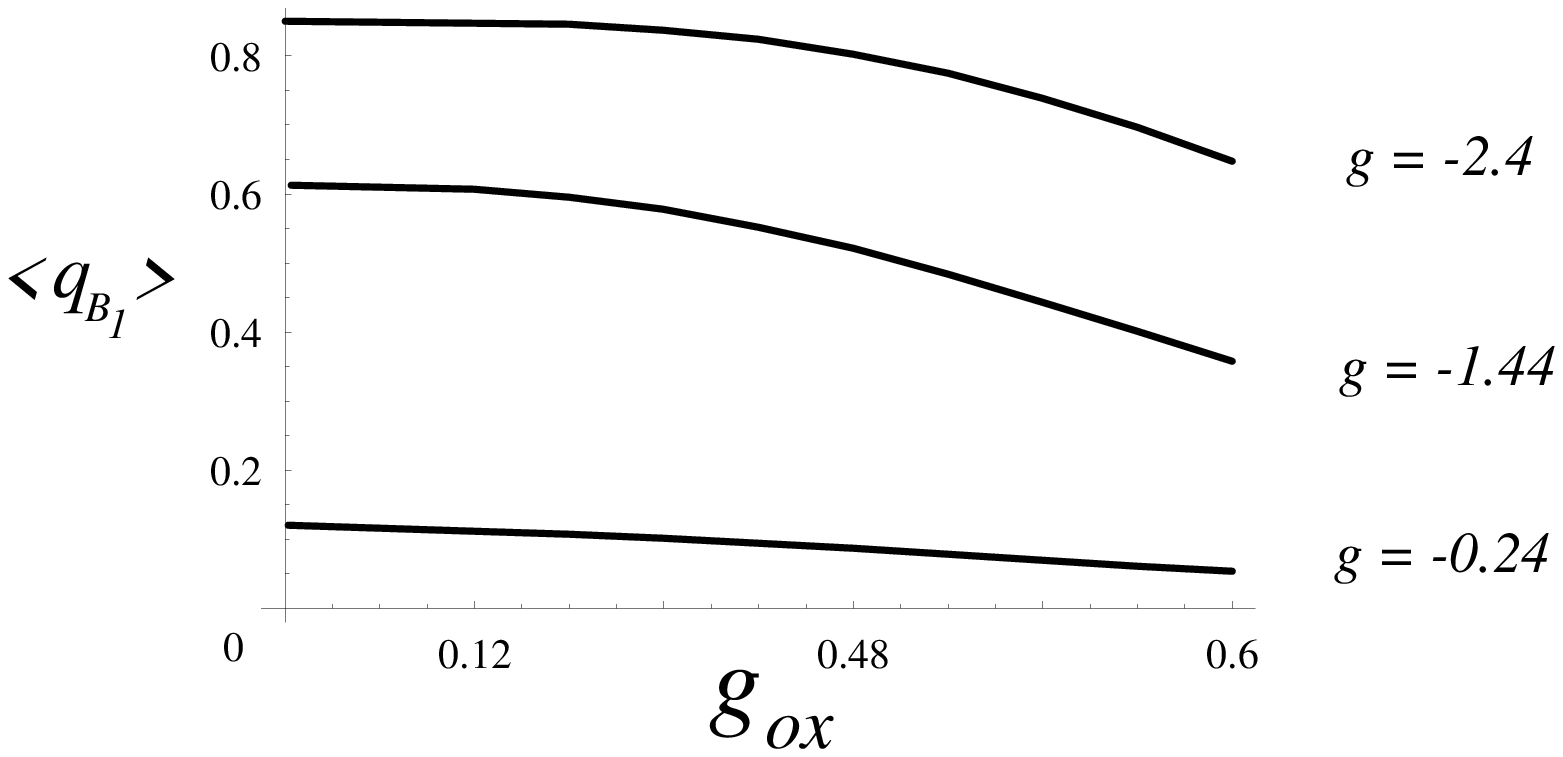,width=7.5cm}
	\caption{\footnotesize{ $\langle \hat{q}_{B_{1}} \rangle$ as a function of
	$- g $ and $ g_{ox}$.
	Here $t=1$eV, $U=5$eV,  $\langle \hat{q}_{B_{1}} \rangle$ is in units of 
	$\xi_{0}$,
	$ g $ and $ g_{ox}$ are in units of  
	$\varepsilon_{0}/\xi_{0}=1$eV$\times \stackrel{\circ}{\rm{A}}^{-1} $}}
	\label{spostb1}
\end{center} 
\end{figure}

\subsection{Four-hole ground state mixing}
With 4 holes, $H_{el}^{{\rm CuO_{4}}}$ has a two-fold degenerate  ground 
state; it belongs to  
the two dimensional irrep  of $S_{4}$ that  breaks 
in $A_{1} \oplus B_{2}$ in $C_{4v}$. Thus, the only  JT-active 
mode is $B_{2}$, which makes the problem exactly resoluble in terms of a continued 
fraction\cite{cida}.

Following again Eqs.(\ref{jtpb}) the second-quantized JT Hamiltonian  with 
four particles reads
\begin{eqnarray}
    H_{JT}(4)= \left[ \hbar \omega_{B_{2}} \, b^{\dagger}_{B_{2}} 
    b_{B_{2}}  + E_{0}(4) 
\right] \otimes {\bf 1}_{2 \times 2} + \nonumber \\
    \x_{B_{2}}\gamma_{5}(b^{\dagger}_{B_{2}}+b_{B_{2}}) \left(
\begin{array}{cc}
 0& 1 \\
 1 & 0
\end{array}
\right) \; .
\label{hjt4h}
\end{eqnarray}
in the space spanned by the electronic ground states.
The coupling constant $\gamma_{5}$ is given by
\begin{equation}
\gamma_{5}= \langle \Psi_{A_{1}} | H_{B_{2}} |\Psi_{B_{2}} \rangle 
= 1.19 \,  g_{ox} \; ,
\label{ga5}
\end{equation}
where as usual, the matrix element in Eq.(\ref{ga5})  is evaluated 
at $U/t \sim 5$.

The ground state energy of $H_{JT}(4)$  
in the sector of symmetry $\eta$ coincides with
the lowest pole of the Green function
\begin{eqnarray}
G_{\eta,\eta}(E)= \langle\langle 0|\otimes \langle \Psi_{\eta}|{1 \over E-H_{JT}(4) +i 0^{+}} 
|\Psi_{\eta}\rangle\otimes|0\rangle\rangle = \nonumber\\
= \frac{1}{\c(0)-\frac{ (\xi_{B_{2}} \gamma_{5})^{2}}{\c(1)
-  \frac{ 3  (\xi_{B_{2}} \gamma_{5})^{2}}{\c(2)  -  
\frac{5 (\xi_{B_{2}} \gamma_{5})^{2}}{\c(3)  -
\frac{7 (\xi_{B_{2}} \gamma_{5})^{2}}{\c(4)- \ldots}}}}} \; 
\end{eqnarray}
where $\eta=A_{1},B_{2}$ and $\c(n) = E-E_{0}(4) - n \hbar \w_{B_{2}}$.
 $G_{\eta,\eta}(E)$ does not
depend by $\eta$ and the energy corrections in the $A_{1}$ and $B_{2}$
sectors are the same. 
The ground state energy shift  $\Delta E$ as a function of $ g_{ox}$ is plotted 
in Fig.\ref{ener4h}.
\begin{figure}[H]
\begin{center}
	\epsfig{figure=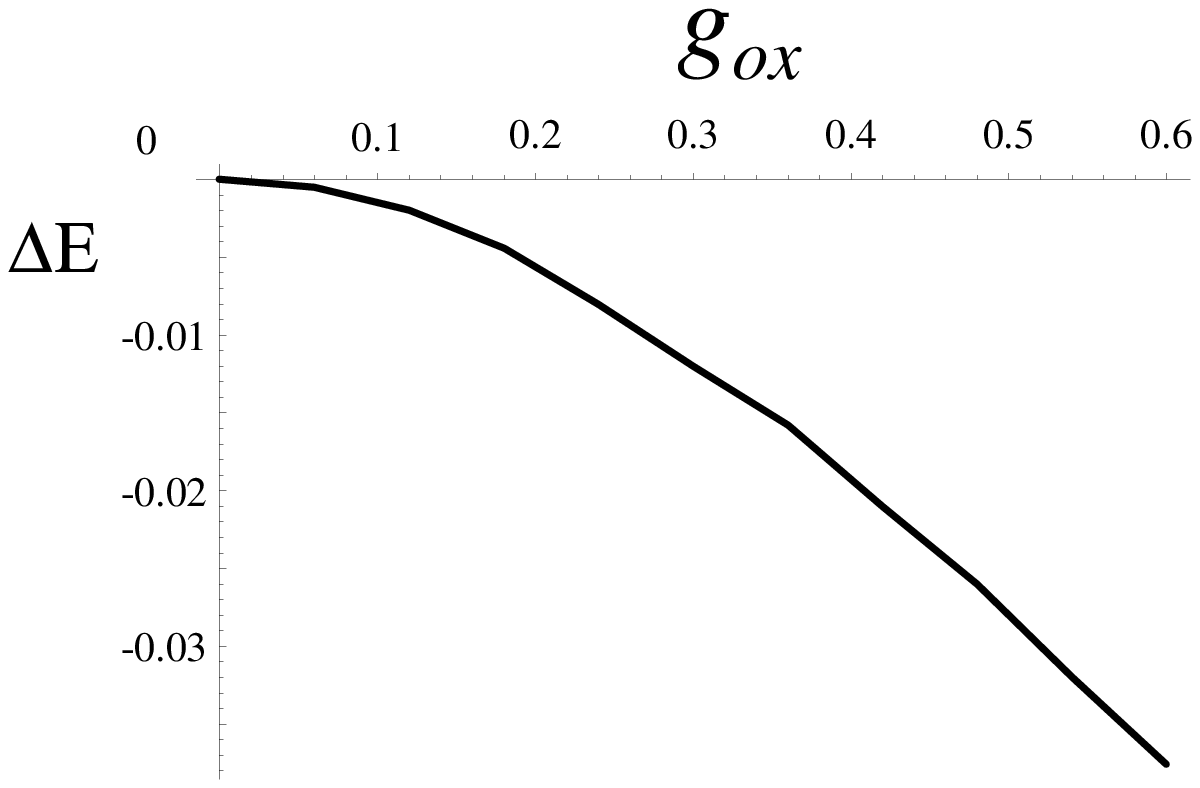,width=5.5cm}
	\caption{\footnotesize{Ground state energy shift  $\Delta E$ 
	as a function of $ g_{ox}$
	Here $t=1$eV, $U=5$eV, $\Delta E$ is in eV, $ g_{ox}$ is in units of  
	$\varepsilon_{0}/\xi_{0}=1$eV$\times 
	\stackrel{\circ}{\rm{A}}^{-1} $, 
	$\lambda_{B_{2}}=1$, $\omega_{B_{2}}=0.1$eV.}}
	\label{ener4h}
\end{center} 
\end{figure}

There are no diagonal couplings in Eq.(\ref{hjt4h}), which implies no 
distorsions; the energy correction is much smaller than in the three holes 
case.

In Fig.\ref{apes4h} we show the adiabatic potential curve, with the 
same parameters as in Fig.\ref{apes4h}. In contrast with 
the three-hole case, the ground state multiplet is   
separated from the excited states  by  $\sim 100 \div 
150$meV which is comparable  the phonon 
energies. Hence we expect that in this case
the approximation restricting the  Hilbert space to the lowest 
multiplet is not justified.  However, in the case of weak coupling to 
soft vibrations, with $\w$ small compared to the gap in the 
electronic spectrum,  this approximation should work well.

\begin{figure}[H]
\begin{center}
	\epsfig{figure=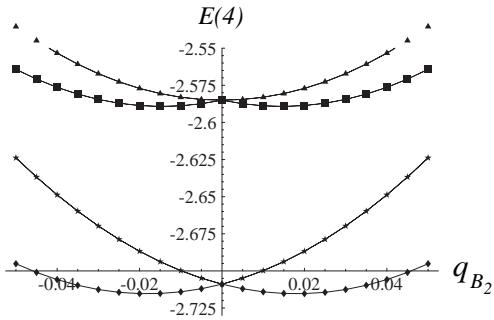,width=6.5cm}
	\caption{\footnotesize{Adiabatic potential energy curve for the low
	lying 4-hole states of $CuO_{4}$.  Here, 
	$q_{B_{2}}$ denotes the classical normal coordinate;  
	$U=5$eV, $t=1$eV, $ g =-2.4$, $ g_{ox}=0.6$ eV/$ 
	\stackrel{\circ}{\rm{A}}$  and 
	$\frac{1}{2} M \omega_{B_{2}}^{2} =20$eV/$ \stackrel{\circ}{\rm{A}} ^{2}$;
	the displacement is in $ \stackrel{\circ}{\rm{A}}$, energies are in eV.
	The ground state multiplet is below the first excited state by 
	$\sim 100 \div 150$meV which is of the order of $2|\tilde{\D}|$.}}
    \label{apes4h}
\end{center} 
\end{figure}

\subsection{$W=0$ pairing in the presence of Jahn-Teller distortions}
\label{jtslega}
Collecting together the  results of the present Section  we obtain the 
behaviour of  $\tilde{\Delta}$ in a popular approximation that neglects the 
excited 
electronic states.
In Fig.\ref{delta} we show the plot of $\tilde{\Delta}$ as a function of
$- g $ and $ g_{ox}$, shown with the zero-energy 
plane. The maximum distorsion along $B_{1}$ 
compatible with  $\tilde{\Delta} <0$ (see Fig.\ref{delta}.b) is  
$ \langle q^{max}_{B_{1}} \rangle \simeq 3 \times 
10^{-2}\stackrel{\circ}{\rm{A}}$,
which is attained  at $ g \simeq 1$eV$\times 
\stackrel{\circ}{\rm{A}}^{-1} $ and $ g_{ox}=0$.
\begin{figure}[H]
\begin{center}
	\epsfig{figure=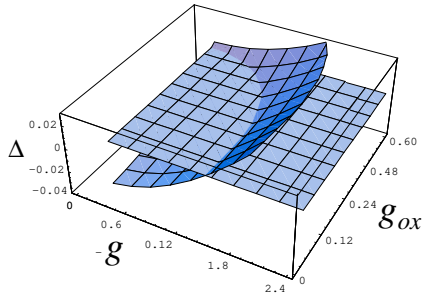,width=5.5cm}
	\caption{\footnotesize{$\tilde{\Delta}$ as a function of
	$- g $ and $ g_{ox}$, according to the theory of the 
	present Section.
	 $\tilde{\Delta}$ is in eV,
	$ g $ and $ g_{ox}$ are in units of  
	$\varepsilon_{0}/\xi_{0}=1$eV$\times \stackrel{\circ}{\rm{A}}^{-1} $
	Here $t=1$eV, $U=5$eV, $\omega_{B_{2}}=0.1$eV.}}
	\label{delta}
\end{center} 
\end{figure}
Both the systems with four and 
three holes gain energy by the JT effect; $\tilde{\Delta}$ remains negative only in the weak 
EP coupling regime, since the decrease of $2E(3)$  overcomes the 
decrease of $E(4)$. This is due to the fact that the system with three
holes can gain energy by  mixing with  $B_{1},B_{2},E_{1x},E_{1y},E_{2x},E_{2y}$ 
vibrations,
while the system with four holes can do it only with the  $B_{2}$ mode. Moreover the factor 2 in 
front of $E(3)$ in the expression for $\tilde{\Delta}$ further favours the 
distorsion in the 3-holes case. 

These results in line with Ref.\cite{mazumdar},  would 
imply that  the electronic pairing is limited to relatively weak EP 
couplings and that at any rate the vibrations are invariably 
detrimental to $ W=0$ pairing. However we know from Sect.\ref{eff}  that 
this conclusion is remarkably but definitely wrong, because the full theory predicts synergy of 
vibrations and $W=0$ pairing at least at weak coupling. This failure of 
the JT Hamiltonian is due to the neglect of the electronic excited states,
causing a severe overestimate of the 4 hole energy. 
The physical reason is that if we restrict the mixing to the 
degenerate states the electronic wave function is too rigid. On the 
other hand, including all the multiplet of states arising from the 
degenerate one-electron level the pair can achieve the flexibility 
which allows it to follow adiabatically the vibration-induced 
deformations,as we shall see in the next Section. 

Detecting the failure of a textbook procedure is by itself a 
potentially very interesting result; it is a merit of a relatively simple model like this 
that allows  to understand in detail how this arises.

\section{Numerical results of the full theory}
\label{num}    
Since the conventional JT Hamiltonian is not enough,
to see what really happens in this model with increasing EP coupling, 
where  the analytic treatment of Sect.\ref{eff} 
loses validity,  we resort to numerical methods.
In this Section we explore the pairing scenario numerically, which 
offers an independent check of the weak-coupling calculations and 
permits to go  beyond the weak 
coupling regime.  First, we analyze one phonon at a time ($\lambda_{\eta} =1$), turning 
off the all others ($\lambda_{\eta} =0$); $\omega_{\eta} \equiv 0.1$ for all 
modes.  In this way we see which kind of 
phonon is cooperative with the $W=0$ pairing and which is not.
We have performed these calculations in a virtually exact way, by 
including a number of phonons $N_{ph}$ up to $20$. To this end we take 
advantage of the recently proposed {\it spin-disentangled} 
diagonalization technique\cite{PRB2003}.
The results are shown in Figs.\ref{soloA1},\ref{B2+E2} and \ref{soloE1}

\begin{figure}[H]
\begin{center}
	\epsfig{figure=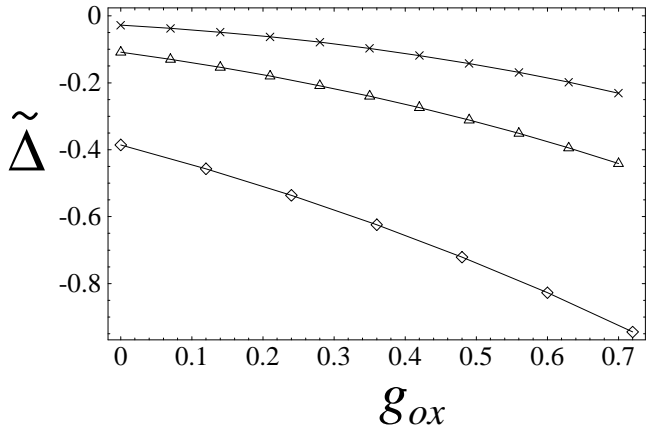,width=8.5cm}
	\caption{\footnotesize{Exact diagonalization results  for ${\tilde 
	\D}(4)$ in eV,
	with only the $A_{1}$ phonon active, $N_{ph}=20$, 
	 as a function of $ g_{ox}$ for different
	values of $ g $: 
	$ g =-0.2$ (crosses);
	$ g =-0.5$ (triangles);
	$ g =-1$ (diamonds).
	Here we used $t=1$eV, $t_{ox}=0$, $U=1$eV, $ g_{ox} $ and
	$ g $  are in  units of  
	$\varepsilon_{0}/\xi_{0}=1$eV$\times \stackrel{\circ}{\rm{A}}^{-1} $
	}}
	\label{soloA1}
\end{center} 
\end{figure}

\begin{figure}[H]
\begin{center}
	\epsfig{figure=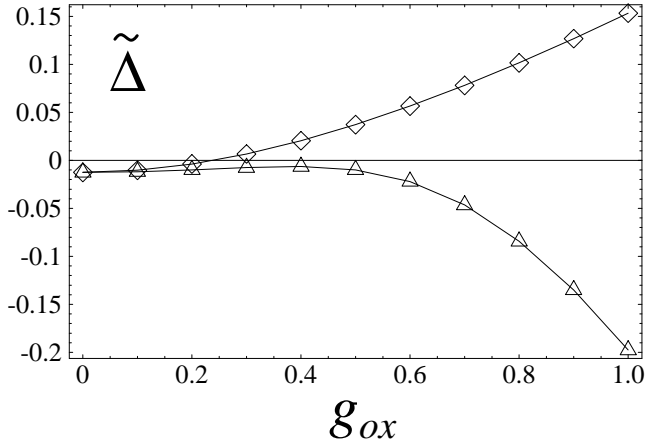,width=8.5cm}
	\caption{\footnotesize{Exact diagonalization results  for ${\tilde 
	\D}(4)$ in eV,
	with only one 
		phonon active, $B_{2}$(triangles) or $E_{2}$ (diamonds), $N_{ph}=20$ .
	Here we used   $t=1$eV,
	$U=1$eV; $ g_{ox} $ and
	$ g $  are in units of  
	$\varepsilon_{0}/\xi_{0}=1$eV$\times \stackrel{\circ}{\rm{A}}^{-1} $
	}}
	\label{B2+E2}
\end{center} 
\end{figure}

\begin{figure}[H]
\begin{center}
	\epsfig{figure=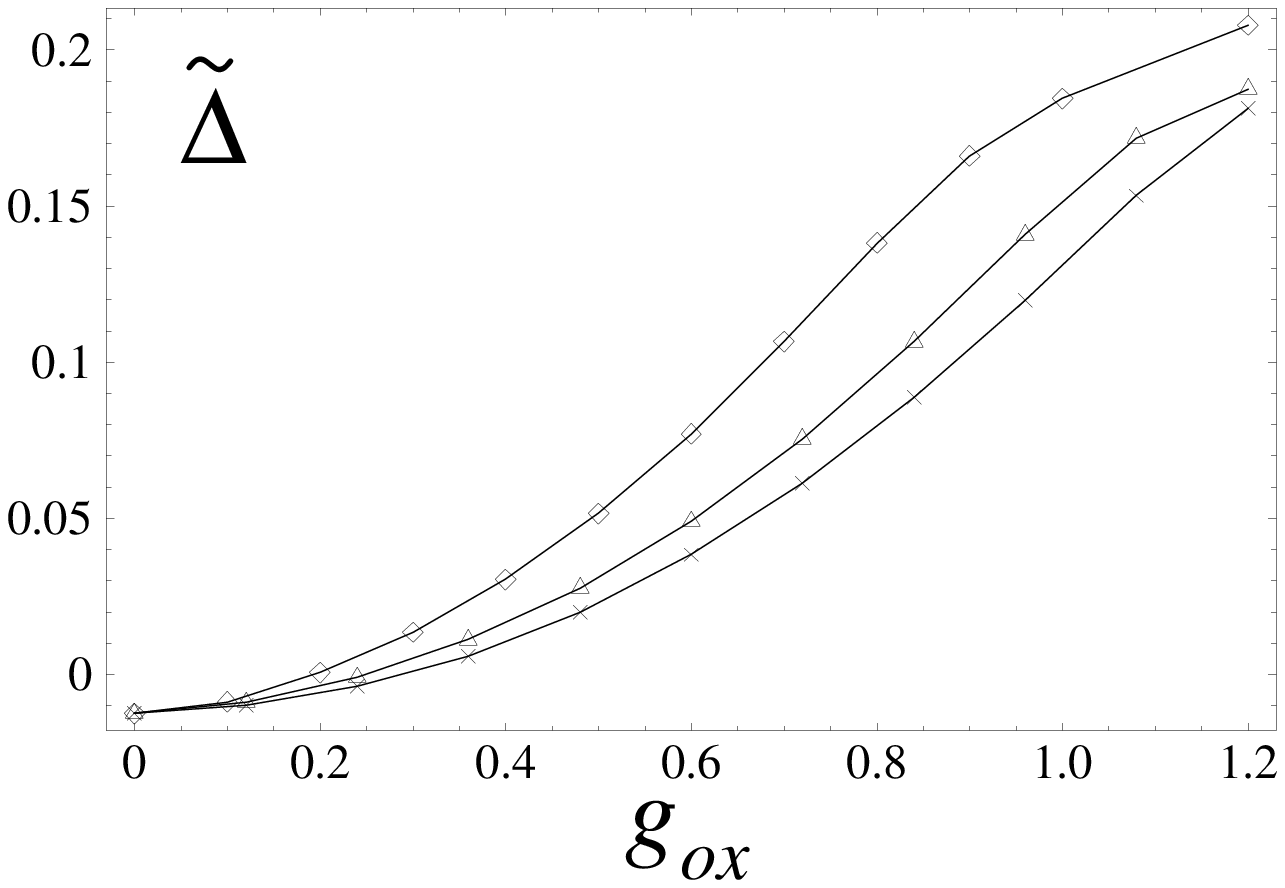,width=8.5cm}
	\caption{\footnotesize{Exact diagonalization results  for ${\tilde 
	\D}(4)$ in eV,
	with only the $E_{1}$ phonon active, $N_{ph}=20$, as a function of 
	$ g_{ox}$ for different
	values of $ g $: 
	$ g =-0.2$ (crosses);
	$ g =-0.5$ (triangles);
	$ g =-1$ (diamonds).
	Here we used $t=1$eV, $t_{ox}=0$, $U=1$eV, $ g_{ox} $ and
	$ g $  are in  units of  
	$\varepsilon_{0}/\xi_{0}=1$eV$\times \stackrel{\circ}{\rm{A}}^{-1} $
	}}
	\label{soloE1}
\end{center} 
\end{figure}

The plots show the trend of ${\tilde \D (4)}$ as a function of 
$ g_{ox}$ and $ g $. It appears (see 
Figs.\ref{soloA1} and \ref{B2+E2}) that if $ g_{ox}$ is 
increased, the $A_{1}$ and  $B_{2}$ 
phonons enhances the pairing , even beyond the weak coupling 
regime.
The further enhancement  of $|{\tilde \D (4)}|$ due to $A_{1}$ as $| g |$ is 
increased is not predicted by the weak coupling theory [Eq.(\ref{phona1})].
The  $B_{1}$ phonon is slightly suppressive, but it affects the pairing energy on a scale 
of $10^{-5}$eV and hence its contribution is negligible.
On the other hand the (longitudinal and transverse) $E$ phonons 
(see Figs.\ref{B2+E2} and \ref{soloE1})
have an unambiguos tendency to destroy the pairing. In particular 
the $E_{1}$ mode does it both by increasing $ g_{ox}$ and   
by increasing $| g |$. 

The presented results show that the behaviour of the individual phonons
is essentially the same as  predicted analytically in Sect.\ref{eff}:
 some of them ($A_{1}$ and $B_{2}$) act in a cooperative way 
with the electronic pairing mechanism; some other ($E$ phonons) does 
not; the $B_{1}$ mode is quite inoperative.
However, for a proper 
understanding of the conflicting vibronic effects we need to
include as many phonon  modes as possible at the same time. 
 In this case the exact diagonalizations  become hard even with a 
 modest 
$N_{ph}$  per vibration. We performed exact diagonalizations of
$H_{el-latt}^{{\rm CuO}_{4}}$ with five active 
modes; with 4 holes,  the size of the 
problem is $100 (N_{ph}+1)^{5}$; we could afford  $N_{ph}=3$ for each.
Some results are shown in Figs.\ref{due}.

In  Fig.\ref{due}.a we included the vibrations with $\eta= A_{1}, 
B_{1}, B_{2}, E_{2x}, E_{2y}$; one notes a strong, monotonic increase of the 
binding energy with both  $ g_{ox}$ and $| g |$. The 
weak-coupling theory of Sect.\ref{eff} qualitatively explains the  $
g_{ox}$ dependence but not the $| g |$ one: when the EP 
coupling gets strong, the Cu-O stretching grows important. In  Fig.\ref{B2+E2}
(diamonds) we noted that the $E_{2}$ vibrations alone tend to 
destroy pairing; here we observe that when they   compete with 
$A_{1}$ and $B_{2}$ their effects  are utterly suppressed.
It is possible that the couplings to the  pair-breaking $E$ modes are 
somewhat  underestimated by the choiche of $\lambda$ parameters.

In 
Fig.\ref{due}.b we included the vibrations with $\eta= A_{1}, 
B_{1}, B_{2}, E_{1x}, E_{1y}$, and we observe that  $\tilde{\Delta}$ 
now becomes positive at moderate $ g_{ox}$. Comparing with  the above results on 
individual modes,  we observe that in going from Fig.\ref{due}.a 
to Fig.\ref{due}.b we are replacing the pair-breaking transversal 
$E_{2}$ phonons  by the   pair-breaking, longitudinal  
$E_{1}$ modes.  We conclude that the longitudinal ones are more 
efficient in restoring the repulsion and at intermediate coupling 
they overwhelm the pair-healing $A_{1}$ and   $B_{2}$.

However for $ g_{ox} \gtrsim 0.15$ eV $\times 
\stackrel{\circ}{\rm{A}}^{-1} $,
the  cooperative  modes win and ${\tilde \D}(4)$ gets negative again. 
This remarkable behavior  could not be 
anticipated by the weak coupling approach of Sect.\ref{eff},
where only the  one-phonon exchange diagrams  were included as in the BCS theory.

In 
Fig.\ref{due}.c the active modes are $\eta=  
B_{1}, E_{1x}, E_{1y}, E_{2x}, E_{2y}$; these are all pair-breaking 
individually and switching them all together, they readily unbind pairs, leading to strong positive 
$\tilde{\D}$. However, unaspectedly,  we  again find 
attraction  at large enough $ g_{ox}$.

It is likely that the couplings to the  pair-breaking $E$ modes are 
somewhat   overestimated by the choiche of $\lambda$ parameters 
 in Fig.\ref{due}.b,c.
The  pairing at strong coupling observed in Figs.\ref{due}.b,c results from more complicated interactions 
leading to bipolaron formation. This recalls  the charge-ordered 
superlattice phase  found in Ref\cite{egami}; 
however they used  a Hubbard-Holstein model and,  since the 
system is one-dimensional,  the electronic pairing does not occour in 
their case.

\begin{figure}%[H]
\begin{center}
	\epsfig{figure=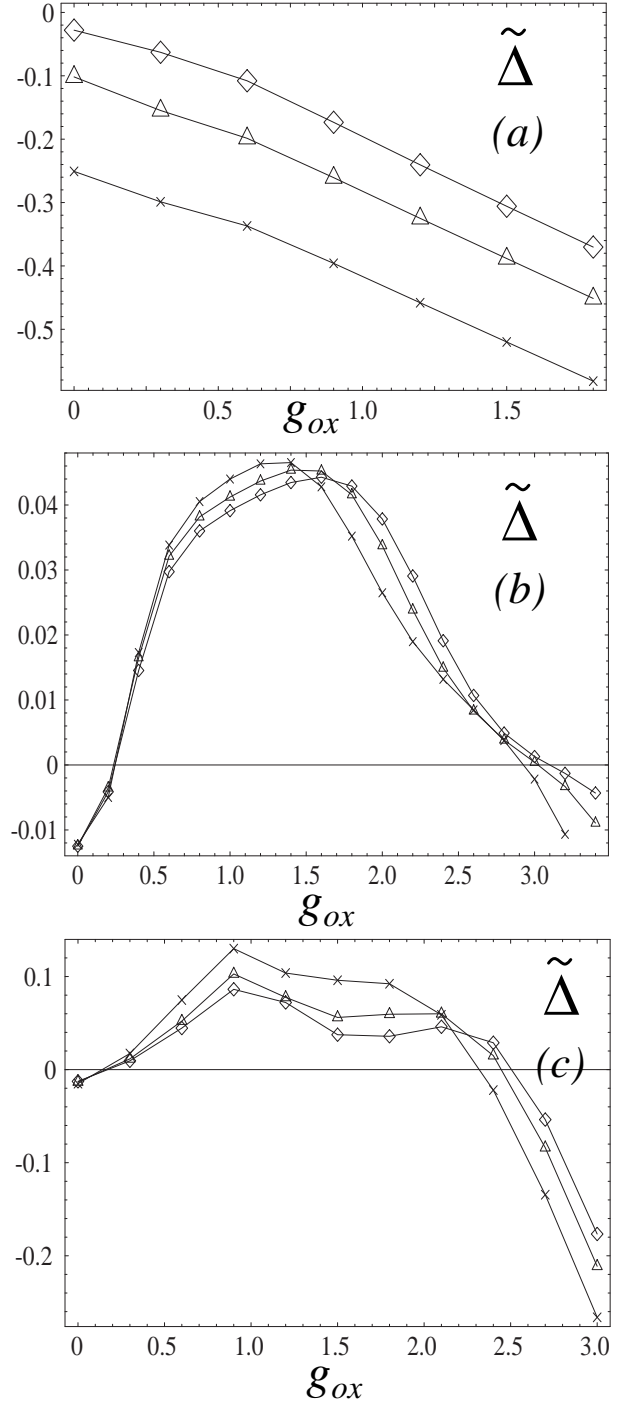,width=8cm}
	\caption{\footnotesize{
	${\tilde \D}(4)$ in eV  as a function of $ g_{ox}$ for different
	values of $ g $.
	  $\lambda_{\eta}=1 $ all the 
	vibrations, except:  $\lambda_{\eta}=0 $ for $\eta= E_{1}$ (a);
  $\lambda_{\eta}=0 $ for $\eta= E_{2}$ (b); $\lambda_{\eta}=0 $ for 
  $\eta=A_{1},B_{2}$ (c).
	$N_{ph}$=3 for each active mode;
	$ g =-0.2$ (diamonds);
	$ g =-0.5$ (triangles);
	$ g =-1$ (crosses).
	Here we used $t=1$eV, $t_{ox}=0$, $U=1$eV, $ g_{ox} $ and
	$ g $  are in  units of  
	$\varepsilon_{0}/\xi_{0}=1$eV$\times 
	\stackrel{\circ}{\rm{A}}^{-1} $. }}
  \label{due}
\end{center} 
\end{figure}

\section{Conclusions}
\label{Conclusions}
 Introducing vibrations and vibronic couplings into a strongly 
 correlated model opens up  a rich scenario where, among other 
 possibilities, pairing can be achieved  by a synergy of electronic correlation 
 and phonon-exchange. A possible outcome, however, is competition  
 among different symmetry vibration modes and electronic excitations. 
 We illustrate the situation by 
using a CuO$_{4}$ model that allows a full treatment of all degrees of 
freedom and hosts  bound $W=0$ pairs when undistorted, has vibrations of 
the same symmetries as the CuO plane and is numerically affordable. 
A popular recipe for computing JT distorted molecules prescribes 
restricting to the degenerate  electronic levels letting them 
interact with the JT active modes. A static treatment 
invariably leads to a complete removal of the symmetry and a nodegenerate 
ground state. We put forth a fuller dynamical theory which partly   
preserves the degeneracy; however, the vibrations are always opposing 
$W=0$ pairing which is thereby reduced to a weak EP coupling effect. 
This restricted basis, however,  may only be valid  provided that the 
excited states of the unperturbed electronic Hamiltonian are far 
removed from the ground state on the energy scale set by the frequency 
of the relevant phonon modes. With the cuprates in mind, we consider 
a situation when the phonon energies  and the superconducting gap are 
comparable, in the 0.1 eV range; we diagonalize the full model keeping 
up to 5 simultaneous modes and   vibrational quantum numbers $N_{ph} 
\leq 3$.
Depending on the parameters, a rich phenomenology emerges from the 
numerical data. 
Pairing prevails at weak EP coupling, as expected, but the phonon 
contributions which dominate in such a case turns out to contribute to the pairing rather than 
opposing it.
The correct trend is predicted by a canonical 
transformation approach, which also explains the pairing or 
pair-breaking character of the modes.
In particular it is found that 
the half-breathing modes give a  synergic contribution to the purely 
electronic pairing; since   they are believed to be   mainly involved in 
optimally doped cuprates, our findings suggest  a joint 
mechanism, with the Hubbard model that captures a crucial part of the physics.

This agreement validates the canonical transformation approach, which 
allows to carry out useful calculations even in large systems that o 
not lend themselves to exact diagonalization. 

At intermediate coupling the outcome of the theory depends essentially 
on the relative weight of the coupling to the longitudinal and 
transverse vector modes, which destabilize pairing most effectively.  

Remarkably, however, the vibrations restore pairing again  at strong  coupling,
when a bipolaronic regime prevails. This scenario was also drawn in the context of an extended t-J model, where  
the half-breathing mode was found to enhance  electronic pairing\cite{ishihara}\cite{scalapino2}. 

Finally,  experimental data on nanopowders\cite{paturi} also indicate that one should not be overly 
pessimistic about  cluster calculations. The pairing that shows up there  can be relevant 
and physically insightful concerning the interplay of various degrees of freedom on 
pair structure and formation.

\section{Appendix A: $W=0$ pairs in the $CuO_{4}$ cluster}

The CuO$_{4}$ Hubbard Hamiltonian has C$_{4v}$ symmetry. When the Oxygen-Oxygen hopping
is absent, the symmetry group is the permutation group $S_{4}$, and 
although for  convenience we continue to  use the subgroup C$_{4v}$ 
labels, it is $S_{4}$ that must be used for the $W=0$ theorem\cite{IJMPB2000}. The character table reads:

\vspace{0.5cm}

\begin{center}

\begin{tabular}{|c|c|c|c|c|c|c|}
   \hline 
$C_{4v}$ & $\mathbf{1}$ & $C_{2}$ & $C^{(+)}_{4},\,C^{(-)}_{4}$ & 
$\sigma_{x},\,\sigma_{y}$ & 
$\sigma_{+},\,\sigma_{-}$ & Symmetry \\
\hline 
$A_{1}$ & 1 & 1 & 1 & 1 & 1 & $x^{2}+y^{2}$\\
\hline 
$A_{2}$ & 1 & 1 & 1 & -1 & -1 & $(x/y)-(y/x)$\\
\hline 
$B_{1}$ & 1 & 1 & -1 & 1 & -1 & $x^{2}-y^{2}$\\
\hline 
$B_{2}$ & 1 & 1 & -1 & -1 & 1 & $xy$ \\
\hline 
$E$ & 2 & -2 & 0 & 0 & 0 & $(x,y)$ \\
\hline 
\end{tabular}
\end{center}
\vspace{0.3cm}
{\footnotesize {\bf Table I:  }{\em Character table of the $C_{4v}$ symmetry group. 
    Here  $\mathbf{1}$ denotes the identity,  $C_{2}$ the  180 degrees 
    rotation,  $C_{4}^{(+)},\;C_{4}^{(-)}$ the 
counterclockwise and clockwise 90 degrees rotations, $\sigma_{x},\;\sigma_{y}$ 
the reflection with respect to the $y=0$ and $x=0$ axis and $\sigma_{+},\;\sigma_{-}$ 
the reflection with respect to the 
$x=y$ and $x=-y$ diagonals.  In the last 
column we show typical basis functions.}}
    \vspace{0.5cm}

Setting for simplicity $\varepsilon_{d}=\varepsilon_{p}=0$, the one-body
spectrum of the CuO$_{4}$ Hamiltonian has the following eigenvalues:
\vspace{0.5cm}
\begin{center}

   \begin{tabular}{|c|c|c|c|c|}\hline 
      ${\varepsilon_{A_{1}}\over t}$ &${\varepsilon_{B_{1}}\over t} $ 
      &${\varepsilon_{E_{x}}\over t}$&
      ${\varepsilon_{E_{y}}\over t}$ &${\varepsilon_{A_{1}'}\over t}$ \\
      \hline
      $\tau  - {\sqrt{4 + {\tau 
 }^2}}$ & $-2\,\tau$ & $0$ & $0$ & $\tau  + {\sqrt{4 + {\tau }^2}}$ \\
 \hline
 -2&0 & 0 & 0& 2\\
 \hline 
\end{tabular}
  
\end{center}
\vspace{0.3cm}
{\footnotesize {\bf Table II: } {\em One-body levels of the CuO$_{4}$ 
cluster in units of $t$; the last line reports the values for 
$t_{ox}=0$ which are used in the text. }}
\vspace{0.5cm}

Here we label the eigenvalues by the irreps of the corresponding 
eigenfunctions.  The level energies are 
in units of $t$, $\tau \equiv t_{ox}/t$;   for 
$t_{ox}=0$, 
$\varepsilon_{E_{x}}=\varepsilon_{E_{y}}=\varepsilon_{B_{1}} =0$. 
The corresponding one body creation operators of a particle in each of these 
eigenstates are:
\begin{eqnarray}
c^{\dagger}_{E_{y} \sigma}& = & \frac{1}{\sqrt{2}} \left( 
p^{\dagger}_{2 \sigma} - p^{\dagger}_{4 \sigma}
\right) \\
c^{\dagger}_{E_{x} \sigma} & = & \frac{1}{\sqrt{2}} \left( 
p^{\dagger}_{1 \sigma} - p^{\dagger}_{3 \sigma}
\right) \\
c^{\dagger}_{B_{1} \sigma} & = & \frac{1}{2} \left( 
p^{\dagger}_{1 \sigma} - p^{\dagger}_{2 \sigma}+p^{\dagger}_{3 \sigma} - p^{\dagger}_{4 \sigma}
\right) \\
c^{\dagger}_{A_{1} \sigma}(1) & = & \frac{1}{\alpha^{2}+4} \left( 
\alpha d^{\dagger}_{\sigma}+p^{\dagger}_{1 \sigma} + p^{\dagger}_{2 \sigma}
+p^{\dagger}_{3 \sigma} + p^{\dagger}_{4 \sigma}
\right) \\
c^{\dagger}_{A_{1} \sigma}(2) & = & \frac{1}{\b ^{2}+4} \left( 
\b d^{\dagger}_{\sigma}+p^{\dagger}_{1 \sigma} + p^{\dagger}_{2 \sigma}
+p^{\dagger}_{3 \sigma} + p^{\dagger}_{4 \sigma}
\right) \\
\end{eqnarray}
where $\alpha$ and $\beta$ depend on $\tau$ as follows:
{\small 
\begin{equation}
 \alpha = \frac{4\,\left( -1 - {\tau }^2 + 
	 \tau \,{\sqrt{4 + {\tau }^2}} \right) }{-5\,
	\tau  - 2\,{\tau }^3 + {\sqrt{4 + {\tau }^2}} + 
       2\,{\tau }^2\,{\sqrt{4 + {\tau }^2}}} \qquad
   \end{equation}
   \begin{equation}
 \beta = \frac{4\,\left( 1 + {\tau }^2 + 
	 \tau \,{\sqrt{4 + {\tau }^2}} \right) }{5\,\tau  + 
       2\,{\tau }^3 + {\sqrt{4 + {\tau }^2}} + 
       2\,{\tau }^2\,{\sqrt{4 + {\tau }^2}}} 
\end{equation}
}
By  the $W=0$ theorem\cite{IJMPB2000}, the irrep $A_{1} \oplus B_{2}$ of the group 
$S_{4}$ which is not represented in the one-body spectrum must yield 
singlet eigenstates with no double occupation.  Projecting  one finds:
\begin{equation}
\begin{array}{lll}
\f ^{\dagger}_{A_{1}} & = & \displaystyle -\frac{2}{\sqrt{3}} 
			c^{\dagger}_{B_{1}\ua}c^{\dagger}_{B_{1}\da} +
			\frac{1}{\sqrt{3}} \left( 
			c^{\dagger}_{E_{x}\ua}c^{\dagger}_{E_{x}\da} +
			c^{\dagger}_{E_{y}\ua}c^{\dagger}_{E_{y}\da}
			\right) \\[10pt]
\f ^{\dagger}_{B_{2}} & = & \displaystyle \frac{1}{\sqrt{2}} \left( 
		c^{\dagger}_{E_{x}\ua}c^{\dagger}_{E_{y}\da} +
		c^{\dagger}_{E_{y}\ua}c^{\dagger}_{E_{x}\da}
			\right) \, . \\
\end{array}
\end{equation}
These are readily verified to create $W=0$ pairs. 

\section{Appendix B:  Pair binding energy}
\label{deltapert}

Here  we calculate $\tilde{\Delta}$ of the $B_{2}$ pair in the 
CuO$_{4}$  
cluster by  second-order perturbation theory 
in both $W$ and $V$  and compare  with $\Delta$  obtained by  solving 
Eqs.(\ref{coopfina1},\ref{coopfinb2}).  Basically the same holds for 
the $A_{1}$ pair.
 $\lambda_{\eta} \equiv 1$ throughout this Appendix.
 
The ground state  with two particles belongs to $A_{1}$ and, using the 
notations of Table III, its energy
reads 
\begin{eqnarray}
&E&(2)=   2 \varepsilon _{A_{1}}
+\frac{5}{16}U+
U^{2} \left[
\frac{3}{128 \, \varepsilon_{A_{1}}} -
\frac{61}{512} d_{1}
\right]- 
\nonumber \\
&g^{2}& \left( 
d_{2} + 2 d_{3}\right)
- g_{ox}^{2} \left( 
d_{3} + d_{4}\right)
+ 2 \sqrt{2}  g_{ox}  g d_{3}.
\label{gs2}
\end{eqnarray}

\begin{center}
    \centering    \vspace{0.5cm}
   \begin{tabular}{|c|c|c|}
     \hline 
$\frac{1}{d_{1}}$ = $-\varepsilon_{A_{1}}+\varepsilon_{A'_{1}}$ & 
$\frac{1}{d_{2}}$ = 
$-\varepsilon_{A_{1}}+\omega_{B_{1}}$ & 
$\frac{1}{d_{3}}$ =
$-\varepsilon_{A_{1}} + \omega_{E_{1}}$ 
\\
\hline 
$\frac{1}{d_{4}}$ = $-\varepsilon_{A_{1}} + \omega_{E_{2}}$ &
$\frac{1}{d_{5}} $ = 
$\varepsilon_{A'_{1}} + \omega_{E_{1}}$ &
$\frac{1}{d_{6}}$ =
$\varepsilon_{A'_{1}} + \omega_{E_{2}}$ 
\\
\hline 
 \end{tabular}
\end{center}
\vspace{0.3cm}
{\footnotesize {\bf Table III: } {\em Shorthand notations used in 
Eqs.(\ref{gs2},\ref{gs3},\ref{gs4}). }}
\vspace{0.3cm}

With 3 particles the ground state is a $E$ doublet, and

\begin{eqnarray}
E(3)=  2 \varepsilon _{A_{1}}
+\frac{7}{16}U+
U^{2} \left[
\frac{5}{64 \, \varepsilon_{A_{1}}} -
\frac{53}{512} d_{1}\right]- 
\nonumber \\ 
 g ^{2} \left[ 
d_{2}+ {3 d_{3} \over 2}  +\frac{d_{5}}{2} 
\right]
- \nonumber \\ 
 g_{ox}^{2} \left[
\frac{2}{ \omega_{B_{2}}} + 
\frac{1}{2 \omega_{E_{1}}} +  \frac{1}{ 2 \omega_{E_{2}}}+ {d_{5} \over 
4}+{d_{6} \over 4}+
\frac{3 d_{3}}{4}  +\frac{3 d_{4}}{4}   \right] -\nonumber \\
   g_{ox}  g  \left[
-{3 \sqrt{2} d_{3} \over 2}+
\frac{\sqrt{2}d_{5}}{2}
\right] \, .
\label{gs3}
\end{eqnarray}
The ground state with 4 particles belongs to $B_{2}$,
as predicted by the canonical tranformation; 

one gets:
\begin{eqnarray}
E(4) =  2 \varepsilon _{A_{1}}+\frac{9}{16}U+
U^{2} \left[
\frac{25}{128 \, \varepsilon_{A_{1}}} -
\frac{29}{512}d_{1}\right]- 
\nonumber \\ 
g^{2} \left[
d_{2}+d_{3}+d_{5}\right] - \nonumber \\
 g_{ox}^{2} \left[
\frac{8}{\omega_{B_{2}}} + \frac{1}{\omega_{E_{1}}} + 
\frac{1}{\omega_{E_{2}}} + 
\frac{d_{3}}{2} + 
\frac{d_{4}}{2}+
\frac{d_{5}}{2} + 
\frac{d_{6}}{2} 
\right] + \nonumber \\
 g_{ox}  g  \sqrt{2}
  \left[
  d_{3} -
  d_{2}
\right] \;.
\label{gs4}
\end{eqnarray}
Finally using Eq.(\ref{tildedelta}) and setting 
$\omega_{\eta}=\omega_{0}$, we  obtain
\begin{eqnarray}
{\tilde \Delta}(4) = -\frac{U^{2}}{16} \left[ 
-\frac{1}{ \varepsilon_{A_{1} }} -\frac{1}{2 ( -\varepsilon_{A_{1}}+\varepsilon_{A'_{1}})}
\right]- g_{ox}^{2} \frac{4}{\omega_{0}} \;.\label{lastlast}
\end{eqnarray}
 This must be compared with Eq.(\ref{coopfinb2}), Sect.\ref{eff}, that 
 can be solved iteratively for  $\Delta(4) =E - 2 \varepsilon_{A_{1}}$ 
 inserting $\varepsilon_{B_{1}}=0$ from Appendix A.  The second 
 iteration yields Eq.(\ref{lastlast}), supporting the identification 
 ${\tilde \Delta}(4)=\Delta(4)$ at this order. 
Both quantities represent  the effective interaction of the dressed 
 $B_{2}$ $W=0$ pair.  Indeed, much 
 information about the ground state energies cancels out if one  applies Eq.(\ref{tildedelta});
 the canonical transformation is a much more 
 practical way to represent the effective interaction.

\section*{Acknowledgements}
We thank Gianluca Stefanucci and Agnese Callegari for useful 
discussions.

}

\end{document}